\documentclass[twocolumn,showpacs,preprintnumbers,amsmath,amssymb,pre,nofootinbib]{revtex4}

\usepackage{graphics}

\newcommand{\ket}[1]{\ensuremath{\left| #1 \right\rangle}}
\newcommand{\br}[1]{\ensuremath{\left\langle #1 \right.}}
\newcommand{\bra}[1]{\ensuremath{\left. \br{#1} \right|}}

\newcommand{\kb}[2]{\ket{{#1}}\bra{{#2}}}

\newcommand{\proj}[1]{\kb{{#1}}{{#1}}}

\newcommand{\trace}[1]{\ensuremath{\mathrm{Tr}\left[{#1}\right]}}
\newcommand{\partrace}[2]{\ensuremath{\mathrm{Tr}_{{#2}}\left[{#1}\right]}}

\newcommand{\mean}[1]{\ensuremath{\left\langle {#1} \right\rangle}}
\newcommand{\func}[1]{\ensuremath{\left[ {#1} \right]}}

\newcommand{\partxy}[2]{\ensuremath{\frac{\partial {#1}}{\partial {#2}}}}

\newcommand{\eqnwrap}{\nonumber \\ &&}

\newcommand{\half}{\ensuremath{\frac{1}{2}}}

\newcommand{\pictlab}[3]
{
\begin{figure}[htb]
    \resizebox{0.5\textwidth}{!}{
        \includegraphics{#1}
        }
    \caption{#2\label{fg:#3}}
\end{figure}
}

\newcommand{\sizepict}[4]
{
\begin{figure}[htb]
  {
     \centering
      \resizebox{#4\textwidth}{!}
      {
        \includegraphics{#1}
      }
      \caption{#2\label{fg:#3}}
  }
\end{figure}
}

\newcounter{assumption}

\begin{document}
\preprint{pi-foundqt-41}
\title{Generalising Landauer's principle}
\author{O. J. E. Maroney}
  \email{omaroney@perimeterinstitute.ca}
\affiliation{The Perimeter Institute for Theoretical Physics\\
31 Caroline Street North, Waterloo\\
Ontario, Canada, N2L 2Y5}
\date{\today}

\begin{abstract}
In a recent paper [Stud. Hist. Phil. Mod. Phys. \textbf{36}, 355
(2005)] it is argued that to properly understand the thermodynamics
of Landauer's principle it is necessary extend the concept of
logical operations to include indeterministic operations.
 Here we examine the
thermodynamics of such operations in more detail, extending the work
of Landauer to include indeterministic operations and to include
logical states with variable entropies, temperatures and mean
energies.  We derive the most general statement of Landauer's
principle and prove its universality, extending considerably the
validity of previous proofs.  This confirms conjectures made that
all logical operations may, in principle, be performed in a
thermodynamically reversible fashion, although logically
irreversible operations would require special, practically rather
difficult, conditions to do so. We demonstrate a physical process
that can perform any computation without work requirements or heat
exchange with the environment. Many widespread statements of
Landauer's principle are shown to be only special cases of our
generalised principle.
\end{abstract}

\maketitle

\section{Introduction}\label{intro}
Landauer's principle holds a special place in the thermodynamics of
computation.  It has been described as ``the basic principle of the
thermodynamics of information processing''\cite{Ben03}.  Yet the
literature on Landauer's principle is focused almost exclusively on
a single, logically irreversible operation and a particular physical
procedure by which this operation is performed\footnote{Although
there little consensus on the naming of these, we will refer to the
logical operation as RESET TO ZERO (or $RTZ$) and the widely used
physical process which embodies this operation will be referred to
as Landauer Erasure (or $LE$).}.

In this paper we seek to analyse the form of Landauer's principle
 in a more general context, building upon the
 consideration of the thermodynamics of indeterministic logical operations\cite{Mar02,Mar05b}.
We will explicitly be considering situations where logical states do
not necessarily have uniform mean energies, entropies
 or even temperatures and we will work in a framework in which logically reversible and irreversible and logically
 deterministic and indeterministic operations can be treated on an equal footing.  Once we have done this we will have
 a single framework in which the different aspects of Landauer's principle can be
 united.  Doing so will help to address criticisms\cite{Shenker2000,Nor05} of the limited validity of previous
 proofs of Landauer's principle, and criticisms\cite{SLGP05} of the
 conclusions of \cite{Mar05b}.

This will lead us to the following generalisation of Landauer's
principle:

\par

\noindent\textbf{Generalised Landauer's principle}
\begin{quote}
A physical implementation of a logical transformation of information
has minimal expectation value of the work requirement given by:
\begin{equation}
\mean{\Delta W} \geq \mean{\Delta E} - T\Delta S
\end{equation}
where $\mean{\Delta E}$ is the change in the mean internal energy of
the information processing system,  $\Delta S$ the change in the
Gibbs-von Neumann entropy of that system and $T$ is the temperature
of the heat bath into which any heat is absorbed.

The equality is reachable, in principle, by any logical
transformation of information, and if the equality is reached the
physical implementation is thermodynamically reversible.
\end{quote}

We start by considering what we mean by a logical state, a logical
operation and the requirements for a physical system to be an
embodiment of such an operation.  We will be considering only the
processing of discrete, classical information here, although we will
be assuming the fundamental physics is quantum\footnote{The analysis
would proceed largely unchanged for classical physics, but it would
be unnecessarily cumbersome to attempt both.  See \cite{Turgut2006}
for a classical treatment.}.

We then construct an explicit physical process, based upon the
familiar ``atom in a box'' model, that implements a generic logical
operation.  Thermodynamically optimising this model will, in
general, require consideration of the probability distribution over
the input logical states.  As this probability distribution is also
required to quantify the Shannon information stored in the system,
we will refer to the combination of the logical operation and the
probability distribution as a logical transformation of information.
The optimal implementation, using the ``atom in a box'' physical
process, shows that the above limit is reachable in principle.  We
then demonstrate that this limit cannot be exceeded by any system
evolving according to a Hamiltonian evolution.

We then consider in more detail the implications of this limit,
including several special cases that correspond to more familiar
expressions of Landauer's principle, when the physical
implementation conforms to a set of conditions which we refer to as
``uniform computing''. We will show a less familiar set of
conditions, which are nevertheless physically possible, which we
call ``adiabatic equilibrium computing'', and which can embody any
logical operation without either exchanging heat with the
environment or requiring work to be performed.  We conclude that any
logical transformation of information can be performed in a
thermodynamically reversible manner.  As this conclusion may seem
surprising, we discuss some of the practical barriers to achieving
this and the particular problems presented by logically irreversible
operations.
\section{Logical States and Operations}\label{s:logic}
Although Landauer's principle is about the thermodynamics of
information processing, very little of the literature surrounding it
attempts to define what is meant by a logical operation and what are
then the minimal requirements of a physical system for it to be
regarded as the embodiment of a logical operation.  Without first
answering this question, it cannot be certain that the most general
relationship between information and thermodynamics has been
discovered.  In this Section the abstract properties of logical
states and operations will be considered.  This leads to constraints
upon a physical system which is required to embody the logical
states and operations.
\subsection{Logical States}
A logical state simply consists of a variable $\alpha$, which takes
a value from a set $\{1,\ldots,n\}$.  If the variable $\alpha$ takes
the value $x$ then this means that the logical proposition
represented by the statement $\alpha=x$ is true.  This paper will
consider only classical information processing on finite machines.
This produces additional properties, whose assumption is usually
implicit\footnote{For analog or quantum information processing some
of these assumptions can be relaxed.  We will not consider the
consequences of this here.}:
\begin{enumerate}
\item The set of values is a finite set (and by implication, discrete).
\item The values are distinct.  In any given instance the variable takes one,
and only one, of the possible values.
\item The values are distinguishable.  In any given instance the value taken by the variable can
be ascertained.
\item The values are stable.  The value taken by the variable cannot change except as a result of
a logical operation.
\end{enumerate}
\subsection{Logical Operations}
A logical operation $LOp$ maps input logical states from the set
$\{\alpha\}$ to output logical states from the set $\{\beta\}$:
\begin{equation}
LOp:\alpha \rightarrow \beta
\end{equation}
The number of input and output states need not be the same.  The
output states from one logical operation may be used as input states
to another logical operation.
\begin{table}[htbp]
  \centering
  \parbox{0.4\textwidth}
  {
    \centering
    \begin{tabular}{c||c}
        \multicolumn{2}{c}{$NOT$} \\ \hline
        IN & OUT \\
        \hline \hline 0 & 1 \\ 1 & 0
    \end{tabular}
    \caption{Logical NOT\label{tb:not}}
  }
  \parbox{0.4\textwidth}
  {
    \centering
    \begin{tabular}{c||c}
        \multicolumn{2}{c}{$RTZ$} \\ \hline
        IN & OUT \\
        \hline \hline 0 & 0 \\ 1 & 0
    \end{tabular}
    \caption{Reset To Zero\label{tb:rtz}}
  }
\end{table}
Tables \ref{tb:not} and \ref{tb:rtz} show the maps for two of the
most commonly\footnote{There is an even more trivial logical
operation: logical Do Nothing $IDN$.  Including this as a logical
operation is not a trivial step, as this is the identity operator!
It must also be included as a time delay operator when one considers
a sequence of logical operations.} encountered logical operations
that act upon two input states $0$ and $1$, the $NOT$ operation and
the Reset To Zero ($RTZ$) operation. These rules can be represented
by:
\begin{eqnarray}
NOT:0 &\rightarrow& 1 \nonumber \\
NOT:1 &\rightarrow& 0 \nonumber \\
RTZ:\{0,1\} &\rightarrow& 0
\end{eqnarray}
where use has been made of the fact that the $RTZ$ operation
transforms both input states into the $0$ output state. The $RTZ$
operation is logically \textit{ irreversible}:
\begin{quote} We
shall call a device logically irreversible if the output of a device
does not uniquely define the inputs. {\raggedleft \cite{Lan61}}
\end{quote}
If multivalued maps(see \cite{Rev1983}[Section 6.1] for example) are
to be considered, it is necessary to also define logically
indeterministic\footnote{In \cite{Mar05b} the term
`non-deterministic' was used.  Unfortunately, this term has a
specific usage in computational complexity classes, which does not
quite correspond to that used here.  To attempt to avoid confusion,
we have changed our terminology to `indeterministic'. In terms of
computational complexity classes, this is closest to probabilistic
computation.  We hope this does not simply introduce more
confusion!} computation:
\begin{quote} We shall call a device logically indeterministic
if the input of a device does not uniquely define the outputs.
\end{quote}
\begin{table}[htbp]
  \centering
  \parbox{0.4\textwidth}
  {
    \centering
    \begin{tabular}{c||c}
        \multicolumn{2}{c}{$UFZ$} \\ \hline
        IN & OUT \\
        \hline \hline 0 & 0 \\
        0 & 1 \\
    \end{tabular}
    \caption{Unset From Zero\label{tb:ufz}}
  }
  \parbox{0.4\textwidth}
  {
    \centering
    \begin{tabular}{c||c}
        \multicolumn{2}{c}{$RND$}\\ \hline
        IN & OUT \\
        \hline \hline 0 & 0 \\ 0 & 1 \\ 1 & 0 \\ 1 & 1
    \end{tabular}
    \caption{Randomisation\label{tb:rand}}
  }
\end{table}
Logically indeterministic operations such as Unset From Zero ($UFZ$)
and Randomise ($RND$) are given in Tables \ref{tb:ufz} and
\ref{tb:rand}, which follow the rules
\begin{eqnarray}
UFZ:0 &\rightarrow& \{0,1\} \nonumber \\
RND:\{0,1\} &\rightarrow& \{0,1\}
\end{eqnarray}
$UFZ$ is logically reversible, while $RND$ is logically
irreversible.

Logically indeterministic operations are perhaps less commonly
encountered than logically deterministic operations, and it has been
questioned whether these are really logical
operations(\cite{SLGP05}, for example, take it as part of the
\textit{ definition} of a logical operation that it be a single
valued map). We include them for a number of reasons:
\begin{enumerate}
\item Most importantly, such operations play a
significant role in the theory of computational complexity classes
for actual computers.  The complexity class $BPP$, (\textit{
Bounded-error Probabilistic Polynomial-time}), represents a class of
computational problems for which the inclusion of logically
indeterministic operations can produce an accurate answer
exponentially faster than any known algorithm consisting only of
logically deterministic operations(see \cite{CN01}[Section 3.2.2],
for example).  Excluding them excludes a genuine class of
computational procedures;
\item By including them we are able to derive a
more coherent general framework for the thermodynamics of
computation.  Excluding them creates an artificial asymmetry and
physical properties ascribed to logically irreversible operations in
the literature may be artefacts of the asymmetry caused by this
exclusion;
\item  Logically indeterministic transformations of information involve the
use of probabilistic inferences.  There is a point of view,
\cite{Boo1854,Jay2003}, that regards probabilistic inferences as a
natural generalisation of deductive logical inferences;
\item Finally, there seems no special reason \textit{not} to include them as
they form a natural counterpart to the concept of logically
irreversible operations.  Any conclusion we can draw that applies to
the set of all such logical operations must necessarily apply to all
logically deterministic operations.  Including logically
indeterministic operations in our analysis will not invalidate its
applicability to logically deterministic operations.
\end{enumerate}
\subsection{Logical Transformation of
Information}\label{ss:logtraninf}
To quantify the information being processed by the logical
operation, the Shannon information measure will be used. This
requires the specification of a probability distribution over the
input and output states. If the logical states input to a
computation occur with probabilities $P(\alpha)$, then the Shannon
information represented by the input states is
\begin{equation}
H_\alpha=-\sum_\alpha P(\alpha)\log_2 P(\alpha)
\end{equation}
During the logical operation these input states are transformed into
output states $\beta$.  When an input state may be transformed into
more than one output state, one must specify the probability
$P(\beta|\alpha)$ for each possible output state.  For logically
deterministic operations, specifying $P(\beta|\alpha)$ is trivial as
$\forall \alpha \  \exists \beta \ P(\beta|\alpha)=1$ (or
equivalently $\forall \alpha \ \exists \beta \  [\forall
\beta^{\prime} \ne \beta \ P(\beta^{\prime}|\alpha)=0]$). Specifying
all the non-zero $P(\beta|\alpha)$ completely specifies the rules of
the logical operation.  We will therefore take the set
$\{P(\beta|\alpha)\}$ as the definition of a general logical
operation.  For logically deterministic operations, this is the list
of all combinations of input and output states that have conditional
probability one, which is simply the truth table for the operation.

After the logical operation, the output states $\beta$ will occur
with probability
\begin{equation}
P(\beta)=\sum_\alpha P(\beta|\alpha) P(\alpha)
\end{equation}
so the Shannon information represented by the output states is
\begin{equation}
H_\beta=-\sum_\beta P(\beta) \log_2 P(\beta)
\end{equation}
When we refer to a logical transformation of information, we will
mean a logical operation, acting upon input states\footnote{For
simplicity, input states for which $P(\alpha)=0$, i.e. which are
certain to not occur, will not be included in this set.}
$\{\alpha\}$, which occur with probabilities $P(\alpha)$, which
transforms the input states to output states $\{\beta\}$ with
conditional probabilities $P(\beta|\alpha)$.

The conditional probability that a given output state $\beta$ was
generated by the input state $\alpha$ is:
\begin{equation}
P(\alpha|\beta) = \frac{P(\alpha) P(\beta|\alpha)}{P(\beta)}
\end{equation}
and the joint probability that there was an input state $\alpha$ and
output state $\beta$ is
\begin{equation}
P(\alpha,\beta)=P(\alpha) P(\beta|\alpha)=P(\beta)P(\alpha|\beta)
\end{equation}
This gives an equivalent formulation of logical determinism and
logical reversibility:
\begin{quote}
A logically deterministic computation is one for which
\begin{equation}
\forall \alpha,\beta \ P(\beta|\alpha) \in \{0,1\}
\end{equation}
A logically reversible computation is one for which
\begin{equation}
\forall \alpha,\beta \ P(\alpha|\beta)\in \{0,1\}
\end{equation}
\end{quote}
This is defined in terms of the set $\{P(\alpha|\beta)\}$.  A
logical operation has been defined only by the set
$\{P(\beta|\alpha)\}$, with the $P(\alpha|\beta)$ dependant upon the
input logical state probabilities $P(\alpha)$.

From
\begin{eqnarray}
P(\alpha|\beta)=0 & \Rightarrow & P(\alpha,\beta)=P(\beta|\alpha)=0 \nonumber \\
P(\alpha|\beta)=1 & \Rightarrow & P(\alpha^{\prime}\ne \alpha|\beta)=0 \nonumber \\
                  & \Rightarrow & P(\alpha^{\prime}\ne \alpha,\beta)=P(\beta|\alpha^{\prime}\ne \alpha)=0
\end{eqnarray}
there is an equivalent definition of logically reversible
computations. An operation is logically reversible, if and only if,
\begin{equation}
\forall \beta  \left[ P(\beta|\alpha) \neq 0 \Rightarrow
\left[\forall \alpha^\prime \neq \alpha \  P(\beta|\alpha^\prime)=0
\right] \right]
\end{equation}
This definition is now independant of the input probability
distribution.

We summarise these properties and some consequences.
\begin{enumerate}
\item \textbf{Logically deterministic operations}
\begin{eqnarray}
\forall \alpha,\beta \  P(\beta|\alpha)& \in &\{0,1\} \nonumber \\
\forall \alpha \left[  P(\alpha|\beta) \neq 0 \right. & \Rightarrow
& \left. \left[ \forall
\beta^\prime \neq \beta \  P(\alpha|\beta^\prime)=0 \right]\right] \nonumber \\
\forall \alpha,\beta \  P(\alpha|\beta) & \in &
\left\{0,\frac{P(\alpha)}{P(\beta)}\right\}
\end{eqnarray}
In the case where a particular $\alpha \rightarrow \beta$ transition
has $P(\beta|\alpha)=1$, we may refer to this as a logically
deterministic {\em transition}, even if the overall operation is not
logically deterministic.
\item \textbf{Logically reversible operations}
\begin{eqnarray}
\forall \alpha,\beta \  P(\alpha|\beta)& \in & \{0,1\} \nonumber \\
\forall \beta \left[ P(\beta|\alpha) \neq 0 \right. & \Rightarrow &
\left. \left[ \forall \alpha^\prime \neq \alpha \
P(\beta|\alpha^\prime)=0 \right]\right] \nonumber \\
\forall \alpha,\beta \  P(\beta|\alpha) & \in &
\left\{0,\frac{P(\beta)}{P(\alpha)}\right\}
\end{eqnarray}
In the case where a particular $\alpha \rightarrow \beta$ transition
has $P(\alpha|\beta)=1$, we may refer to this as a logically
reversible {\em transition}, even if the overall operation is not
logically reversible.
\end{enumerate}
\subsection{Physical Representation of Logical
States}\label{ss:physrepstates} We will now consider what the
properties above imply for the physical embodiment of logical states
and operations upon them.  The physical system will have a state
space of possible microstates $\{\mu\}$.  How can these be used to
embody the logical states?
\begin{enumerate}
\item  A particular logical state $\alpha$ will be identified with a set of microstates $\{\mu_\alpha\}$ in the state
space, in the sense that when the physical state of the system is
one of the microstate $\mu \in \{\mu_\alpha\}$, then the logical
state takes the value $\alpha$.
\item As logical states are distinct, a given microstate can be identified with one, and only one, input state.
Each set of microstates $\{\mu_\alpha\}$ is therefore
non-intersecting with any other such set of microstates:
\begin{equation}
\{\mu_\alpha\} \bigcap \{\mu_{\alpha^{\prime} \ne \alpha}\} =
\emptyset
\end{equation}
\item For the logical states to be distinguishable, it is necessary that it is possible to ascertain the set
to which the microstate belongs.  We are not considering analogue
information processing, so the physical interactions must not need
to be sensitive to arbitrarily close (using a natural distance
measure) states in state space.  We replace the point in state space
$\mu$ with the neighbourhood of that point $R(\mu)$. The logical
state $\alpha$ is now identified with the region of state space
corresponding to the union of all the neighbourhoods
$\{R(\mu_\alpha)\}$.  The neighbourhoods corresponding to different
logical states must be non-overlapping.
\item We can now identify the proposition for the logical state $\alpha$ with the projector $K_\alpha$ onto the region of
state space $\{R(\mu_\alpha)\}$
\begin{eqnarray}
K_\alpha K_{\alpha^{\prime}}&=&\delta_{\alpha\alpha^{\prime}} K_\alpha \nonumber \\
\sum_\alpha K_\alpha &=& I \nonumber \\
K_\alpha \func{R(\mu_\alpha)} &=& R(\mu_\alpha) \nonumber \\
K_\alpha \func{R(\mu_{\alpha^{\prime} \ne \alpha})} &=& 0
\end{eqnarray}
The proposition $\alpha$ is true if the state $\mu$ is in the region
of state space $\{R(\mu_\alpha)\}$ projected out by $K_\alpha$.
\item For the physical representation of the logical states to be complete, then it must also be the case that if the
state $\mu$ is in the region of state space $\{R(\mu_\alpha)\}$
projected out by $K_\alpha$, then the logical proposition
corresponding to the logical state $\alpha$ is true.
\item For the logical states to be stable, then under the normal evolution of the system, a microstate
within the region of state space corresponding to a given logical
state must stay within that region of state space\footnote{A weaker
condition, acceptable for most practical needs, is that the
probability of the microstate leaving the region of a given logical
state, during the time scale of the information processing, must be
very low.}. The normal evolution of the system is, trivially, a
physical embodiment of the `logical Do Nothing $IDN$' operation.
\end{enumerate}
\subsection{Physical Representation of Logical Operations}\label{ss:phyreplogop}
During the normal evolution of a system, logical states do not
change.  To perform non-trivial logical operations new interactions
must alter the evolution of the state space.  All the essential
characteristics of a logical operation are included in the set
$\{P(\beta|\alpha)\}$.   It follows that a physical process is an
embodiment of a logical operation if, and only if, the evolution of
the microstates in the physical process are such that, over an
ensemble of microstates in the region $\{R(\mu_\alpha)\}$, the
probability that the microstate ends up in the region
$\{R(\mu_\beta)\}$ is just $P(\beta|\alpha)$.
\begin{enumerate}
\item We will assume that the laws of physics are Hamiltonian.  The evolution of microstates over the state space
of the combined system of the logical processing apparatus and the
environment must be described by a Hamiltonian evolution operator.
\item If the interaction of the microstates of the system and the environment are such that any individual
microstate $\mu_\alpha$ starting in state $\alpha$ is randomised so
that it ends up in the output state $\beta$ with probability
$P(\beta|\alpha)$ then we do not need to be sensitive to the initial
probability distribution of the ensemble of microstates within the
logical state $\alpha$.  In general, however, we may need to be
sensitive to the initial probability distribution $\rho_\alpha$ over
microstates corresponding to the logical state $\alpha$.

\item The complete statistical state of the logical processing system input to the logical operation is
\begin{equation}
\sum_\alpha P(\alpha) \rho_\alpha
\end{equation}
where
\begin{equation}\forall \alpha, K_\alpha \rho_\alpha K_\alpha = \rho_\alpha
\end{equation}
\item The complete statistical state of the logical processing system output from
the logical operation is required to be
\begin{equation}
\sum_\beta P(\beta) \rho_\beta
\end{equation}
where
\begin{equation} \forall \beta, K_\beta  \rho_\beta K_\beta = \rho_\beta
\end{equation}
\end{enumerate}
We have not considered separate systems for the logical input
states, the logical processing apparatus or the output states.  At
first, this seems to assume that the system embodying the logical
input states must be the same as the system embodying the logical
output states and that the logical processing apparatus cannot have
internal states - which would seem to be quite a strong restriction.
This is not the case.  Let us consider the case where there are
three distinct systems: the input state system, with states
$\{\rho_\alpha\}$; an output state system, with states
$\{\rho_\beta\}$; and an auxiliary system corresponding to all
internal and external components of the process, with states
$\{\rho^{App}\}$.

The statistical state is described at the start of the operation by
\begin{equation}
\sum_\alpha P(\alpha) \rho_\alpha \otimes \sum_\beta f_\beta
\rho^{App}_\beta \otimes \rho_\beta
\end{equation}
where we have assumed that the input state system is initially
uncorrelated to the apparatus but have not assumed the output state
system is initially uncorrelated to the internal states
$\{\rho^{App}_\beta \}$ of the apparatus.  The effect of the
operation would be to evolve the combined system into some new
correlated state, combining the three systems:
\begin{equation}
\sum_{\alpha,\beta} P(\alpha,\beta) \rho^{\prime}_{\alpha,\beta}
\otimes \rho^{App}_{\alpha,\beta} \otimes \rho_\beta
\end{equation}

Our approach here is then to consider the state space of the combined
system of input, output and apparatus as a single state space, with
input states for $\alpha$ of:
\begin{equation}
\rho_\alpha \otimes \left( \sum w_\beta \rho^{App}_\beta \otimes
\rho_\beta \right)
\end{equation}
and output states for $\beta$ of:
\begin{equation}
\left(\sum_\alpha P(\alpha,\beta) \rho^{\prime}_{\alpha,\beta}
\otimes \rho^{App}_{\alpha,\beta} \right) \otimes \rho_\beta
\end{equation}

We then consider a Hamiltonian evolution on the combined state space
to be the operation.  This cleanly separates the logical states,
embodied by the physical state of the combined state space, from the
logical operation, embodied by the Hamiltonian evolution on that
state space.  We have not restricted ourselves by the assumption of
a Hamiltonian evolution on a single state space, as we have the full
generality of all possible Hamiltonian interactions allowed between
the input state system, the output state system and the logical
processing apparatus.  We have avoided, on the other hand, any need
to consider the restrictions and complications that would arise if
we constructed models based upon specific assumptions as to how the
input state, output state and logical processing apparatus systems
are allowed to interact.  This completes the physical
characterisation of logical states and operations.
\subsection{Logical vs Microscopic Determinism and Reversibility}
There is one final issue that needs to be stated, for the sake of
clarity, regarding the (absence of a) relationship between logical
and microscopic indeterminism and irreversibility.
\begin{quote}
\textbf{Logical indeterminism} does not imply or require the
existence of any fundamental indeterminism in the microscopic
dynamics of the physical states.  Neither is \textbf{logical
determinism} incompatible with the existence of fundamental
indeterministic dynamics.
\end{quote}
A specific microstate from the input logical state may evolve
deterministically into a specific microstate of an output logical
state, while the operation remains logically indeterministic,
provided the set of the input microstates corresponding to the same
input logical state do not all evolve, with certainty, into
microstates of the same output logical state.

A specific microstate from an input logical state may evolve
indeterministically into a number of possible microstates, while the
operation remains logically deterministic, provided that the set of
the input microstates corresponding to the same input logical state
can only evolve into microstates from the set corresponding to the
same output logical state.
\begin{quote}
\textbf{Logical irreversibility} does not imply or require the
existence of any fundamental irreversibility in the microscopic
dynamics of the physical states.  Neither is \textbf{logical
reversibility} incompatible with the existence of fundamental
irreversible dynamics.
\end{quote}
A specific microstate from the input logical state may evolve
reversibly into a specific microstate of an output logical state,
while the operation remains logically irreversible, provided the set
of the microstates corresponding to the same output logical state
have not all evolved, with certainty, from microstates of the same
input logical state.

A specific microstate from an input logical state may evolve
irreversibly into a specific microstate, while the operation remains
logically reversible, provided that the set of microstates
corresponding to the same output logical state can only have evolved
from microstates in the set corresponding to the same input logical
state.
\section{Thermodynamics of Logical Operations}\label{s:theproof}
We now undertake the main task of this paper: to determine the
limiting thermodynamic cost to a logical operation.  We will do this
in two steps.

Firstly, we will construct a physical process, capable of
implementing any logical operation, as we have defined them, and we
will consider the optimum thermodynamic cost to the process. This
optimum will be considered in two ways: for individual transitions
between specific logical states; and as an expectation value over an
ensemble of operations.  Both work required to perform the process
and heat generated by the process will be calculated, where it is
assumed that all heat generated is absorbed by a heat bath at some
reference temperature $T_R$.

To calculate the expectation values, we must consider the
probability distribution over the input logical states. For this we
will use the probability distribution used to calculate the Shannon
information being processed.  The optimum process will, of
necessity, involve various idealisations (such as frictionless
motion and quasi-static processes) that cannot be achieved in
practice.  The purpose is to demonstrate not that it is possible to
build such optimal operations, but rather that there is no physical
limitation, in principle, on \textit{ how close} one can get to
them.

Then we will prove that there cannot exist any physical process that
can implement the same logical transformation of information, but
with a lower expectation value for either the work requirement or
the heat generation.  The optimum process, for our particular
implementation of a logical transformation of information, is also
the optimum for any possible implementation of that transformation.

\subsection{Statistical mechanical assumptions}
We will now clearly state the statistical mechanical assumptions that are being made.  There are a number of different approaches to the foundations of statistical mechanics and, as the models discussed here involve such idealisations as the treatment of individual atoms, it is important to be clear which approach is being taken.  In this article we will assume the standard structure of Gibbs canonical statistical mechanics: we will be dealing with Hamiltonian flows with probability distributions over a state space, we will assume that a system that has been thermalised can be represented by a canonical distribution over its accessible state space, and will be initially statistically independent of any other system.  While these assumptions are clearly open to debate, a full discussion or justification of them lies outside the scope of this article (although see \cite{Maroney2007a}).

\begin{enumerate}
\item The system consists of the logical processing apparatus (including auxiliary systems as discussed in Section \ref{ss:phyreplogop}) and a number of heat baths.  A heat bath is simply a system that has been allowed to thermalise at some temperature and is sufficiently large that any energy transfer with the logical processing apparatus will have negligible effect upon the heat baths internal energy.  The Hamiltonian for the combined system is:
\begin{equation}
H=H_L+\sum_i (H_i+V_i)
\end{equation}
where $H_L$ is the internal Hamiltonian for the logical processing apparatus, $H_i$ the internal Hamiltonian of the heat bath $i$ and $V_i$ is the interaction Hamiltonian between the logical processing apparatus and the heat bath $i$.  We assume there is no interaction between heat baths.  The density matrix of the combined system is $\rho_C$, and $\rho_L$ is the marginal density matrix after tracing over the heat bath subsystems.
\item Work is performed upon the apparatus through the variation of some
externally controlled parameter $X$, which affects the energy
eigenvalues and eigenstates\footnote{More generally, one should consider a number of
controllable parameters, which are each varying in time.}.
\begin{equation}
H_L(X)=\sum_n E_n(X) \proj{E_n(X)}
\end{equation}
The mean
work performed, as the parameter is varied from $X_0$ to $X_1$ is
given by:
\begin{equation}
\Delta W=\int_{X_0}^{X_1} \trace{\partxy{H_L(X)}{X}\rho_L(X)} dX
\end{equation}
Note that the density matrix $\rho_L$ may be varying as $X$ varies.  We will assume that neither the internal Hamiltonians of the heat baths nor the interaction Hamiltonians have controllable parameters: work is only performed upon the logical processing apparatus itself.

\item The mean change in internal energy of the logical processing apparatus is \begin{equation} \Delta
E=\trace{H_L(X_1)\rho_L(X_1)}-\trace{H_L(X_0)\rho_L(X_0)}
\end{equation}

\item  If we now assume negligible changes in interaction energies:
\begin{equation}
\forall i, \trace{V_i \rho_C(X_1)} \approx \trace{V_i \rho_C(X_0)}  
\end{equation}
then
\begin{equation}
\Delta W - \Delta E=\sum_i \Delta Q_i
\end{equation}
where 
\begin{equation}
\Delta Q_i=\trace{H_i\rho_i(X_1)}-\trace{H_i\rho_i(X_0)}
\end{equation} is the increase in internal energy of the heat bath $i$, and $\rho_i$ is the marginal density matrix of the heat bath, after tracing over the logical processing apparatus and all other heat baths..
\item \label{list:sofar} If the evolution of density matrix is such that it always remains
diagonalised by the energy eigenstate basis, so that
\begin{equation} \rho_L(X)=\sum_n p_n(X) \proj{E_n(X)}
\end{equation}then:
\begin{eqnarray}
\Delta W&=&\int_{X_0}^{X_1} \sum_n p_n(X) \partxy{E_n(X)}{X} dX\\
\sum_i \Delta Q_i &=&\int_{X_0}^{X_1} \sum_n \partxy{p_n(X)}{X} E_n(X) dX
\end{eqnarray}

It is important to note that these \ref{list:sofar} points make no assumption regarding the identification of either thermodynamic entropy or thermal distributions.  Neither the canonical distribution nor the Gibbs-von Neumann entropy has been used.

The following results depend upon the assumption that a heat bath is represented by a canonical distribution and that a limiting ideal case exists of thermalisation through a succession of brief interactions with small subsystems of a heat bath.  The calculations are well known (see \cite{Gib1902,Tol1938, Maroney2007a}, for example) and the results are stated here purely for clarity.  No formal identification of the Gibbs-von Neumann entropy with thermodynamic entropy is required to derive these results.

\item A system that is brought into contact with an ideal heat bath
will, over time periods long with respect to its thermal relaxation
time, be well represented by a canonical probability distribution
\begin{equation}
\rho_\alpha=\frac{e^{-H/kT}}{\trace{e^{-H/kT}}}
\end{equation} over accessible states of the system, with the $T$ being the temperature
of the heat bath, and $H$ the Hamiltonian of the system over the accessible subspace.
\item In the limit of isothermal quasistatic processes, the system is in contact with an
ideal heat bath at some temperature, and system stays in thermal
equilibrium with the heat bath at all times.
\item In the limit of adiabatic quasistatic processes (or
essentially isolated\cite{Tol1938} processes) the system always
remains in a (canonically distributed) thermal state but there is
zero mean energy flow out of the system ($\Delta W=\Delta E$).  The
temperature of this state may vary.
\item We will assume that the only systems with which the information processing system
interacts are ideal heat baths at temperatures
$\{T_\alpha\},\{T_\beta\}$ and $T_R$, and a work reservoir, and that there are no initial correlations between the system and the
heat baths.
\end{enumerate}
While these assumptions involve significant idealisations, they are
the kind of idealisations that are standard in thermodynamics and
statistical mechanics.  Rather than representing a physically
achievable process, they represent the limit of what can be
physically achieved.  There is no physical reason why one cannot, in
principle, get arbitrarily close to these results.

Although the value of the Gibbs-von Neumann entropy, $-k \trace{\rho
\ln \func{\rho}}$ will be calculated for the input and output logical states, all
results in this Section, in terms of work required and heat generated, are
derivable, from the assumptions stated, without needing
to identify this property with thermodynamic entropy\footnote{See
\cite{Maroney2007a} and in \cite{Mar02}[Chapter 6] where this kind
of calculation is carried out in detail for same the kinds of
systems considered here.}.

\subsection{Generic logical operation}\label{ss:genlogop}

\subsubsection{Input logical states}
We start the operation with the logical states represented by
physical states with the properties:
\begin{enumerate}
\item An input logical state, $\alpha$, to the logical computation is physically embodied by a system confined
to some region of state space.  The distribution over the
microstates of that region gives the density matrix $\rho_\alpha$.
\item $\rho_\alpha$ has mean energy $E_\alpha=\trace{H_L\rho_\alpha}=\trace{H_\alpha \rho_\alpha}$, where $H_\alpha=K_\alpha H_L K_\alpha$.
\item For simplicity, in the main section, we will assume the input logical state $\alpha$ is canonically distributed, as if it has been thermalised with a heat bath at temperature $T_\alpha$. \begin{equation}
\rho_\alpha=\frac{e^{-H_\alpha/kT_\alpha}}{\trace{e^{-H_\alpha/kT_\alpha}}}
\end{equation}

This assumption is not essential, and can easily be relaxed without affecting any result.  If the initial density matrix is not a canonical distribution, then it is possible to construct a unitary operator that acts upon the system in isolation and rotates it into a canonical state, with neither heat nor work requirement.  

To give an explicit construction, suppose the initial Hamiltonian and density matrix are $H^{(i)}_\alpha$ and $\rho^{(i)}_\alpha$, such that 
\begin{equation}
\rho^{(i)}_\alpha \neq \frac{e^{-H^{(i)}_\alpha/kT_\alpha}}{\trace{e^{-H^{(i)}_\alpha/kT_\alpha}}}
\end{equation}
for any $T_\alpha$.  Given the diagonal representation
\begin{equation}
\rho^{(i)}_\alpha=\sum_n p_n \proj{\lambda_n}
\end{equation}
then the Hamiltonian, acting between $0<t<\tau$,
\begin{eqnarray}
H_{A}&=& \left[\cos^2\left(\frac{\pi t}{2 \tau}\right)-
\sin^2\left(\frac{\pi t}{\tau}\right)\right] H^{(i)}_\alpha \eqnwrap
+\left[\sin^2\left(\frac{\pi t}{2 \tau}\right)-
\sin^2\left(\frac{\pi t}{\tau}\right)\right] H_\alpha \eqnwrap -\frac{2
\imath \hbar}{\tau} \sin^2\left(\frac{\pi t}{\tau}\right) \ln
\func{\sum_n \kb{\gamma_n}{\lambda_n}}
\end{eqnarray}
with
\begin{eqnarray}
H_\alpha&=&\sum_n E_n \proj{\gamma_n} \\
E_n&=&E_\alpha-kT_\alpha \left(\ln(p_n)-\sum_m p_m \ln(p_m) \right)
\end{eqnarray}
varies continuously from $H^{(i)}_\alpha$ to $H_\alpha$, and has the effect of leaving the system, after time $\tau$, in the stationary canonical
state
\begin{equation}
\rho_\alpha=\frac{e^{-H_\alpha/kT_\alpha}}{\trace{e^{-H_\alpha/kT_\alpha}}}=\sum_n p_n
\proj{\gamma_n}
\end{equation}
$\rho_\alpha$ is unitarily equivalent to $\rho^{(i)}_\alpha$ and $\trace{H_\alpha \rho_\alpha}=\trace{H^{(i)}_\alpha \rho^{(i)}_\alpha}$.  The mean work requirement is zero and no heat is exchanged with the environment.  It should be noted that this construction holds even if $\rho^{(i)}_\alpha$ is not diagonalised in the eigenstates of $H^{(i)}_\alpha$.

\item The Gibbs-von Neumann entropy of the input logical state is:
$S_\alpha=-k \trace{\rho_\alpha \ln \func{\rho_\alpha}}$.
\item There are $M$ possible input logical states.
\end{enumerate}
\subsubsection{Output logical states} The output logical states may be similarly characterised by:
\begin{enumerate}
\item An output logical state, $\beta$, from the logical computation is physically embodied by a system confined
to some region of state space.  The distribution over the
microstates of that region gives the density matrix $\rho_\beta$.
\item $\rho_\beta$ has mean energy $E_\beta=\trace{H^\prime_L \rho_\beta}=\trace{H^\prime_\beta \rho_\beta}$, where $H^\prime_\beta = K_\beta H^\prime_L K_\beta$.
\item Again, for convenience we will assume the output logical state $\beta$ is canonically distributed as if it has been thermalised at a temperature
$T_\beta$.  This will make the density
matrix:
\begin{equation}
\rho_\beta=\frac{e^{-H^\prime_\beta/kT_\beta}}{\trace{e^{-H^\prime_\beta/kT_\beta}}}
\end{equation}
Again, this assumption is easily dropped.  If the final state is required to be a non-canonical density matrix $\rho^{(f)}_\beta$, with Hamiltonian $H^{(f)}_\beta$, then $\rho^{(f)}_\beta$ can be obtained from $\rho_\beta$ by constructing $H^\prime_\beta$ and $H_{B}$ in the same manner as $H_\alpha$ and $H_A$ above.
\item The Gibbs-von Neumann entropy of the output logical state is
$S_\beta=-k \trace{\rho_\beta \ln \func{\rho_\beta}}$.
\item There are $N$ possible output logical states.
\item The output logical state $\beta$ must occur with probability $P(\beta|\alpha)$ given input logical state $\alpha$.
\end{enumerate}

We will also note here that probabilities enter the calculation at two levels: as a probability distribution over the microstates within a given logical state, and as a probability distribution over the different logical states.  We will attempt to keep these formally separate.  From this point onwards, the microstate probability will be represented only by the density matrix.  Explicitly appearing probabilities and averages will always refer to the probability distribution over the logical states.

\subsection{The transformation of information.}
If we consider the actual microstate of the system, within the
region corresponding to a given logical state, as being some
free\footnote{Not externally controlled.} parameter, then the
physical representation of each logical state may initially be
regarded as a potential well with some arbitrary shape, such that
the free parameter is confined within the well.  The potential wells
associated with different logical states are in different regions of
physical space, separated by high potential barriers, such that
there is a very low possibility of transitions between different
logical states.

This can be represented as an atom in one of a number of boxes,
where the free parameter is the location of the atom in the box. The
logical state is represented by the particular box, or potential
well, within which the atom is confined.  The physical
transformation of the information will take place in nine steps.
Steps 1 through to 3 will bring the input logical states into
standardised physical states at a shared reference temperature.
Steps 4 is the logically indeterministic implementation of the
$P(\beta|\alpha)$ transition. Step 5 and Step 6 implement the
joining together of the $\beta$ output states from the different
$\alpha$ input states, giving the logically irreversible stage.
Steps 6 through to 9 then alter each output logical state to the
required final physical state.

Calculations for work requirements, heat generation and so forth,
follow the statistical mechanical calculations above.  Particularly
detailed calculations for `atom in a box' type systems are
considered in references such as \cite{Zur84,Lef95,BBM00,Mar02}. The
key results can be summarised. The Hamiltonian for an infinite
square well potential, of width $l$, holding an atom of mass $m$ is
\begin{equation}
H(l)=\sum_n \frac{\hbar^2 \pi^2}{8ml^2}n^2 \proj{E_n}
\end{equation}
Work is performed upon the system by varying the $l$ parameter
(width of the box).

In a canonical thermal state at temperature $T$, the mean energy is
\begin{equation}
E=\frac{\sum_n \frac{\hbar^2 \pi^2}{8ml^2}n^2
e^{-\frac{\hbar^2 \pi^2}{8ml^2kT}n^2}}{\sum_n e^{-\frac{\hbar^2
\pi^2}{8ml^2kT}n^2}} \approx \frac{1}{2}kT
\end{equation}
and the Gibbs-von Neumann entropy is
\begin{eqnarray}
S&=&\frac{\sum_n \frac{\hbar^2 \pi^2}{8ml^2T}n^2 e^{-\frac{\hbar^2
\pi^2}{8ml^2kT}n^2}}{\sum_n e^{-\frac{\hbar^2 \pi^2}{8ml^2kT}n^2}}
+k \ln \func{\sum_n e^{-\frac{\hbar^2 \pi^2}{8ml^2kT}n^2}} \nonumber
\\
 &\approx & \frac{k}{2}\ln \func{\frac{2 e m k T l^2}{\pi \hbar^2}}
\end{eqnarray}
The approximations hold when the temperature is high with respect to
the ground state:
\begin{equation}
kT \gg \frac{\pi \hbar^2}{2 e m}
\end{equation}

\begin{enumerate}
\item The first step is to continuously and slowly deform the potential well of each separate logical state into
a square well potential.

The square well should be deformed to width $d^{(1)}_\alpha$
\begin{equation}
d^{(1)}_\alpha=\left( \sqrt{\frac{\pi \hbar^2}{2 e m k T_\alpha}}
\right) e^{S_\alpha / k}
\end{equation}
where $m$ is the mass of atom.

This state has mean energy and entropy
\begin{eqnarray}
E^{(1)}_\alpha &=& \half k T_\alpha \nonumber \\
S^{(1)}_\alpha &=& S_\alpha
\end{eqnarray}
If this deformation is carried out sufficiently slowly, the mean
heat generation is zero and the work requirement is
\begin{eqnarray}
W^{(1)}_\alpha &=& \half k T_\alpha-E_\alpha
\end{eqnarray}
This is a mean work requirement for the operation.  Fluctuations may
occur around this value.

The system is may now be pictured as a box, divided with $M-1$
partitions.  When the atom is located between the $\alpha-1$ and
$\alpha$ partitions, the system is in logical state $\alpha$.  This
can be seen in Figure \ref{fg:p20f3a}(a).

\begin{figure}[htb]
    \resizebox{0.5\textwidth}{!}{\includegraphics{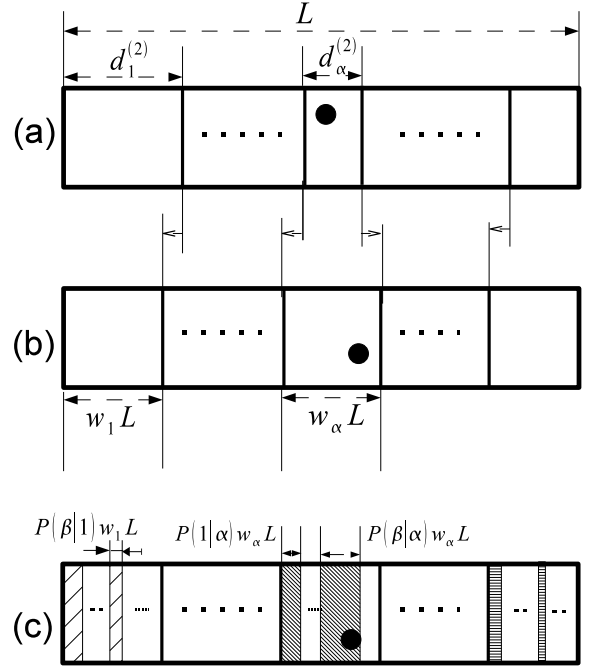}}
    \caption{Arranging input states\label{fg:p20f3a}}
\end{figure}


\item Remove the system from all contact with heat baths and then, slowly,
adiabatically vary the width of each square well to
$d^{(2)}_\alpha$:
\begin{equation}
d^{(2)}_\alpha=d^{(1)}_\alpha \sqrt{\frac{T_\alpha}{T_R}}
\end{equation}
At the limit of a slow, quasistatic process, this will leave each
logical state with a density matrix equal to a canonical thermal
system with temperature $T_R$.  Mean energy, entropy and mean work
requirements are:
\begin{eqnarray}
E^{(2)}_\alpha &=& \half k T_R \nonumber \\
S^{(2)}_\alpha &=&  k \ln \func{d^{(2)}_\alpha \left(
\sqrt{\frac{2 e m k T_R}{\pi \hbar^2}} \right)}=S_\alpha  \nonumber \\
W^{(2)}_\alpha &=& \half k T_R - \half k T_\alpha
\end{eqnarray}

As the total width of the box is now
\begin{equation}
L=\sum_{\alpha^\prime} d^{(2)}_{\alpha^\prime} =\left(
\sqrt{\frac{\pi \hbar^2}{2 e m k T_R}} \right) \sum_{\alpha^\prime}
e^{S_{\alpha^\prime} / k}
\end{equation}
then
\begin{equation}
d^{(2)}_\alpha =L
\frac{e^{S_\alpha/k}}{\sum_{\alpha^\prime}e^{S_{\alpha^{\prime}}/k}}
\end{equation}

\item Now bring the entire system into contact with
heat baths at the reference temperature $T_R$.  Slowly and
isothermally move the positions of the potential barriers separating
the square wells(see Figure \ref{fg:p20f3a} (a-b) ). Move the
$i^{th}$ barrier to the position $x_i$:
\begin{equation}
x_i= L \sum_{\alpha=1}^{\alpha=i} w_\alpha
\end{equation}
where $\sum_\alpha w_\alpha=1$.  The values of $w_\alpha$ have not
been specified. Varying these will be used to optimise the
operation.

Each logical state now has a width $d^{(3)}_\alpha=w_\alpha L$. If
$w_\alpha=0$ for one of the input states $\alpha$ this stage will
compress the volume of that state to zero.  Clearly this can only be
allowed to take place if there is no possibility that the partition
is occupied by the atom!

\begin{eqnarray}
E^{(3)}_\alpha &=& \half k T_R \nonumber \\
S^{(3)}_\alpha &=& k \ln \func{d^{(3)}_\alpha \left(
\sqrt{\frac{2 e m k T_R}{\pi \hbar^2}} \right)} \nonumber \\
W^{(3)}_\alpha &=& k T_R \ln
\func{\frac{d^{(3)}_\alpha}{d^{(2)}_\alpha}}  \nonumber \\
Q^{(3)}_\alpha &=& k T_R \ln
\func{\frac{d^{(3)}_\alpha}{d^{(2)}_\alpha}}
\end{eqnarray}
where  $Q^{(3)}_\alpha$ is the heat generated in the heat bath if
the atom is in the $\alpha$ partition.

\sizepict{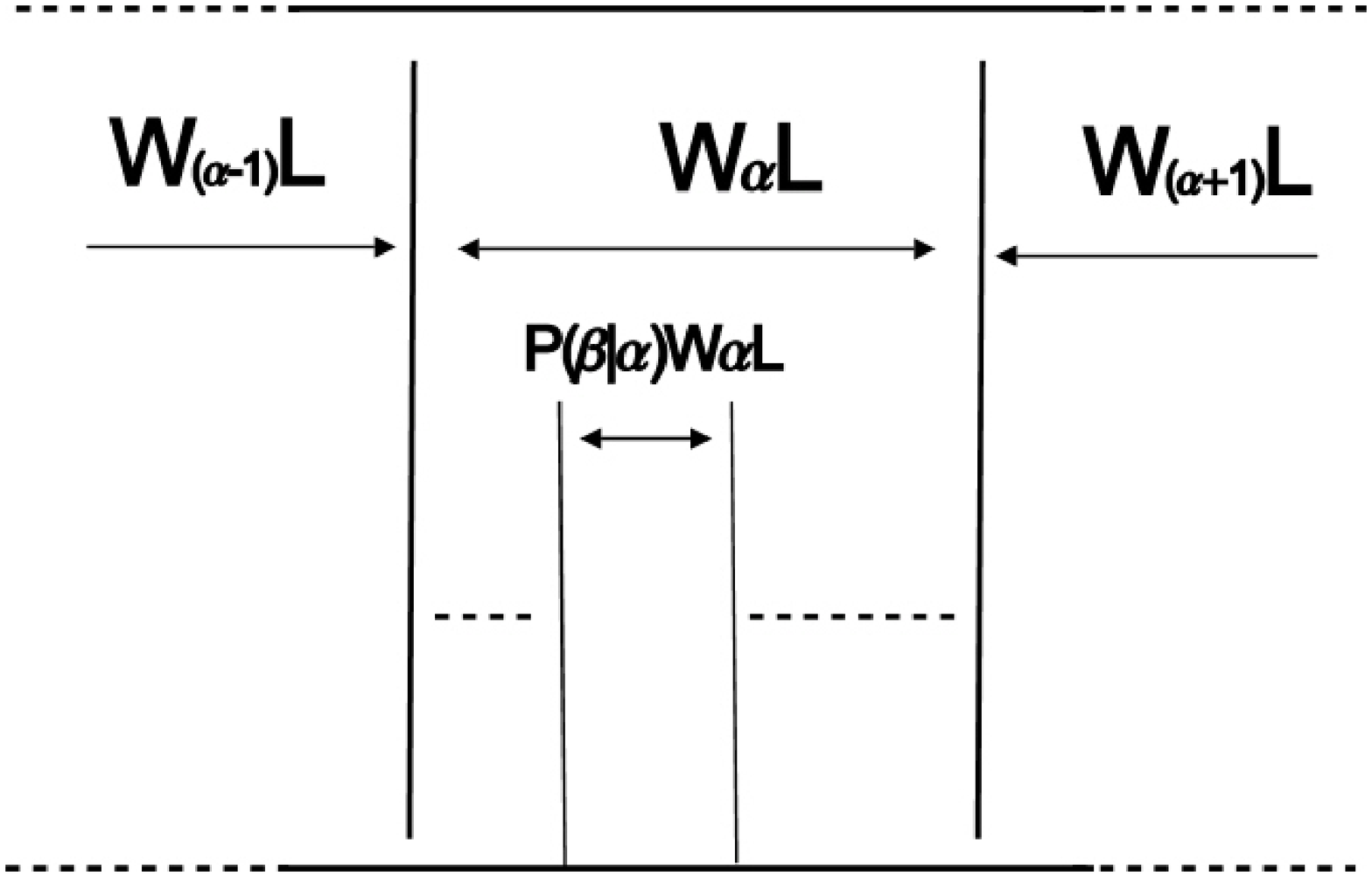}{Inserting subpartitions}{p12f5}{0.3}

\item Insert $N-1$ new potential barriers slowly into each partition.  Within a given
partition $\alpha$, the barriers should be spaced according to the
probabilities $P(\beta|\alpha)$ of the logical operation (see Figure
\ref{fg:p12f5}).  They have a width
\begin{equation}
d^{(4)}_{\alpha,\beta}=P(\beta|\alpha)w_\alpha L
\end{equation}

 This is the
logically indeterministic step of the computation, in Figure
\ref{fg:p20f3a} (b-c).  There are $M$ partitions, each with $N$
subpartitions.

 \pictlab{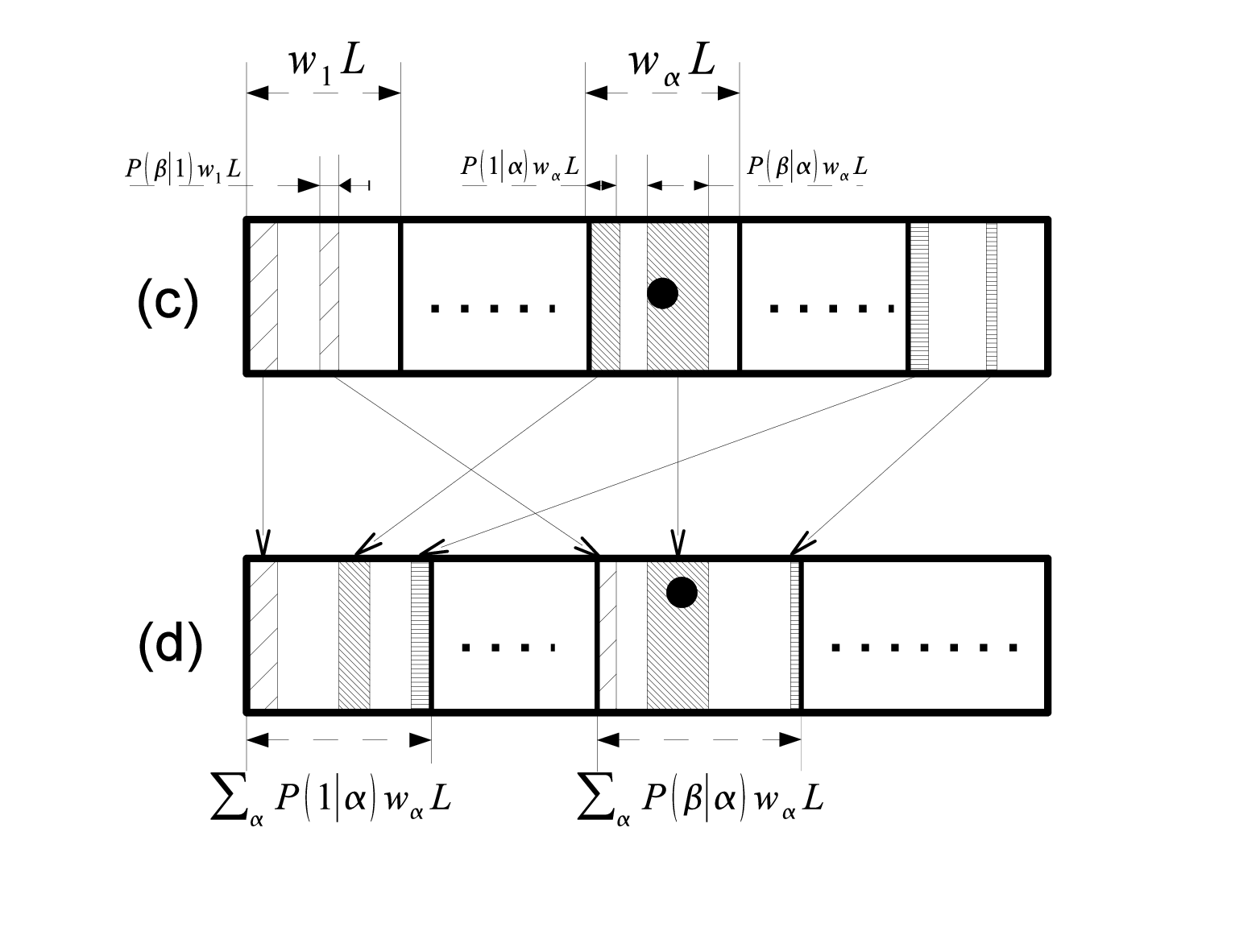}{Rearranging the
partitions}{p20f5a}

If the atom was located in partition $\alpha$ beforehand, and the
system has been allowed to thermalise at temperature $T_R$ for a
period of time greater than the thermal relaxation time, then the
probability of the atom now being located in the
$\left(\alpha,\beta\right)$ subpartition is $P(\beta|\alpha)$.  For
logically deterministic operations, then all non-zero
$P(\beta|\alpha)$ are equal to one and no partitions need be
inserted.

The mean energy, and the entropy, associated with the atom being
located within a particular $\left(\alpha,\beta\right)$ subpartition
is:
\begin{eqnarray}
E^{(4)}_{\alpha,\beta} &=& \half k T_R \nonumber \\
S^{(4)}_{\alpha,\beta} &=& k \ln \func{d^{(4)}_{\alpha,\beta} \left(
\sqrt{\frac{2 e m k T_R}{\pi \hbar^2}} \right)}
\end{eqnarray}

\item Now rearrange the
subpartitions so that, for each $\beta$, all the $\beta$ output
partitions are adjacent. From each $\alpha$ partition, we gather
first subpartition, corresponding to output logical state $\beta=1$,
and collect them together.  Repeat this for each set of $\beta$
subpartitions, from all the $\alpha$ partitions.  Finally this
produces a sequence of $N$ $\beta$ partitions, each with $M$
$\alpha$ subpartitions. This is illustrated in Figure
\ref{fg:p20f5a}.

\pictlab{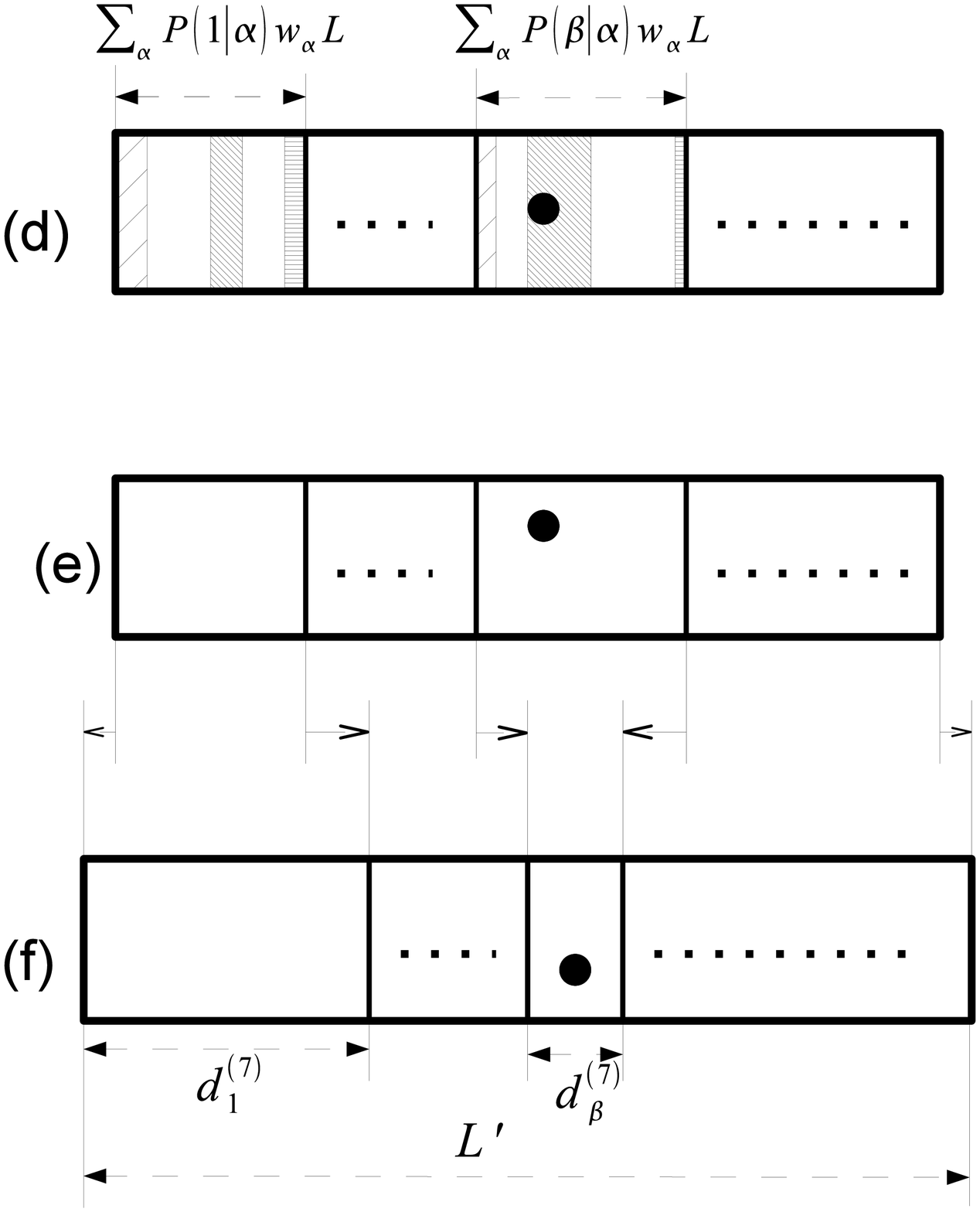}{Arranging the output states}{p20f6a}

\item Remove the potential barriers within each $\beta$ output
partition and leave the box for a time that is long in comparison to
the atoms thermal relaxation time.

The $\beta$ partition has a width
\begin{equation}
d^{(6)}_\beta=w_\beta L
\end{equation}
where
\begin{equation}
w_\beta = \sum_\alpha w_\alpha P(\beta|\alpha)
\end{equation}
and if the atom is located in the $\beta$ partition, then
\begin{eqnarray}
E^{(6)}_\alpha &=& \half k T_R \nonumber \\
S^{(6)}_\alpha &=& k \ln \func{d^{(6)}_\beta \left( \sqrt{\frac{2 e
m k T_R}{\pi \hbar^2}} \right)}
\end{eqnarray}

This is the logically irreversible stage and is illustrated in
Figure \ref{fg:p20f6a} (d-e).  This stage is trivial for logically
reversible computations, for which each $\beta$ output partition is
composed of only one $\alpha$ subpartition, and so has no internal
barriers. Note also that if $\forall \alpha, P(\beta|\alpha)=0$,
then the atom can never be located in the $\beta$ partition.

\item
Now slowly and isothermally resize the output partitions. The
barriers should be moved until the $\beta$ partition has width
\begin{equation}
d^{(7)}_\beta=\left( \sqrt{\frac{\pi \hbar^2}{2 e m k T_R}} \right)
e^{S_\beta / k}
\end{equation}
See Figure \ref{fg:p20f6a} (e) to (f).

The overall width of the box may change by this operation, and is
now:
\begin{equation}
L^{\prime}=\sum_{\beta^\prime} d^{(7)}_{\beta^\prime}=L
\frac{\sum_{\beta^\prime}\exp^{S_{\beta^{\prime}}/k}}{\sum_{\alpha^\prime}\exp^{S_{\alpha^{\prime}}/k}}
\end{equation}
so that
\begin{equation}
d^{(7)}_\beta=L^\prime \frac{e^{S_\beta /
k}}{\sum_{\beta^\prime}\exp^{S_{\beta^{\prime}}/k}}
\end{equation}

For the atom located in the $\beta$ partition, we have:
\begin{eqnarray}
E^{(7)}_\beta &=& \half k T_R \nonumber \\
S^{(7)}_\beta &=& k \ln \func{d^{(7)}_\beta \left(
\sqrt{\frac{2 e m k T_R}{\pi \hbar^2}} \right)}=S_\beta  \nonumber \\
W^{(7)}_\beta &=& k T_R \ln
\func{\frac{d^{(7)}_\beta}{d^{(6)}_\beta}}  \nonumber \\
Q^{(7)}_\beta &=& k T_R \ln
\func{\frac{d^{(7)}_\beta}{d^{(6)}_\beta}}
\end{eqnarray}
where $Q^{(7)}_\beta$ is the heat generated in the heat bath.

\item Now remove all contact from the $T_R$ heat baths.  With the
system thermally isolated, slowly and adiabatic resize the output
partitions to the widths:
\begin{equation}
d^{(8)}_\beta=d^{(7)}_\beta\sqrt{\frac{T_R}{T_\beta}}
\end{equation}
If the atom is in the $\beta$ partition, the effect of this
quasistatic, adiabatic evolution is to leave the atom in a canonical
thermal state with temperature $T_\beta$.

\begin{eqnarray}
E^{(8)}_\beta &=& \half k T_\beta \nonumber \\
S^{(8)}_\beta &=& S_\beta  \nonumber \\
W^{(8)}_\beta &=& \half k T_\beta - \half k T_R
\end{eqnarray}

\item The output logical states $\beta$ are now all at the required temperature,
and entropy.  For completeness, bring each separate $\beta$
partition into thermal contact with a heat bath at the appropriate
temperature $T_\beta$ and slowly, continuously and isothermally
deform the shape of each square well potential into the final
potential for the output logical state.

\begin{eqnarray}
E^{(9)}_\beta &=& E_\beta \nonumber \\
S^{(9)}_\beta &=& S_\beta  \nonumber \\
W^{(9)}_\beta &=& E_\beta - \half k T_\beta
\end{eqnarray}
\end{enumerate}
This completes the physical implementation of the logical operation.
\subsection{Thermodynamic costs} The procedure detailed in the
previous Section fulfils the requirements of a generic logical
operation.  The input logical states are represented by the
appropriate physical input states, the output logical states are
represented by the appropriate physical output states, and the
transitions between them occur with probabilities $P(\beta|\alpha)$.
\subsubsection{Individual transitions}
Adding up the work and heat values, across all steps, for a system
which starts in logical state $\alpha$ and ends in logical state
$\beta$ gives:

\begin{eqnarray}
 \Delta W_{\alpha,\beta} & = & \left(E_\beta -T_R S_\beta\right)
 -\left(E_\alpha -T_R S_\alpha\right) +k T_R \ln
 \func{\frac{w_\beta}{w_\alpha}} \nonumber \\
 \Delta Q_{\alpha,\beta} & = & T_R \left(S_\alpha-S_\beta+ k\ln
 \func{\frac{w_\beta}{w_\alpha}}\right)
 \end{eqnarray}
\begin{enumerate}
\item For a logically reversible transition,
\begin{equation}\frac{w_\beta}{w_\alpha}=P(\beta|\alpha)\end{equation}
 and so is independant
of the choice of $w_\alpha$.  If the transition is also logically
deterministic, $P(\beta|\alpha)=1$ and the logarithmic term is zero.
The work requirements are
\begin{equation}
 \Delta W_{\alpha,\beta}= \left(E_\beta -T_R S_\beta\right)
 -\left(E_\alpha -T_R S_\alpha\right)
 \end{equation}
 \item If the logically reversible transition is {\em indeterministic}, the
work requirement is {\em reduced} by the quantity $-k T_R \ln
\func{P(\beta|\alpha)}$.  If $P(\beta|\alpha)$ is small, this term
can be large, even to the extent of making the work requirement
negative (i.e. implying work may be {\em extracted} from the
process).
\item Now consider logically irreversible transitions.  When the
transition is logically deterministic, $w_\beta$ is the sum of all
the $w_\alpha$ values where the transition is permitted.  It is
therefore always the case that $\frac{w_\beta}{w_\alpha} \geq 1$.
This implies an {\em increased} work requirement compared to a
logically reversible, deterministic transition between equivalent
$(\alpha,\beta)$ states.
\item Finally logically irreversible, indeterministic transitions may,
in principle, take values for $\frac{w_\beta}{w_\alpha}$  both above
and below $1$.
\end{enumerate}
Let us consider optimising the thermodynamic cost of an individual
$\alpha \rightarrow \beta$ transition.  The only free variables are
the $w_\alpha$.  For logically reversible transitions, these have no
effect and the cost is always:
\begin{eqnarray}
 \Delta W_{\alpha,\beta} & = & \left(E_\beta -T_R S_\beta\right)
 -\left(E_\alpha -T_R S_\alpha\right) +k T_R \ln
 \func{P(\beta|\alpha)} \nonumber \\
 \Delta Q_{\alpha,\beta} & = & T_R \left(S_\alpha-S_\beta+ k\ln
 \func{P(\beta|\alpha)}\right)
 \end{eqnarray}

For logically irreversible transitions, the quantity
$\frac{w_\beta}{w_\alpha}$ should be made as small as possible,
subject to the constraint that $\sum_\alpha w_\alpha=1$.  From
$w_\beta=\sum_{\alpha^\prime} w_{\alpha^\prime}
P(\beta|\alpha^\prime)$ it must be the case that
\begin{equation}
w_\beta \geq w_{\alpha} P(\beta|\alpha)
\end{equation}

Equality is reached by setting $w_{\alpha^\prime}=0$, for all the
input logical states $\alpha^\prime \neq \alpha$ where
$P(\beta|\alpha^\prime) \neq 0$.  This gives $w_\beta=w_\alpha
P(\beta|\alpha)$.  If the transition is a logically deterministic
one, $\frac{w_\beta}{w_\alpha}=1$, otherwise
$\frac{w_\beta}{w_\alpha} < 1$ and the work requirement is reduced
(as for a logically reversible, indeterministic transition).  The
result is similar to logically reversible transitions:
\begin{eqnarray}
 \Delta W_{\alpha,\beta} & \geq & \left(E_\beta -T_R S_\beta\right)
 -\left(E_\alpha -T_R S_\alpha\right) +k T_R \ln
 \func{P(\beta|\alpha)} \nonumber \\
 \Delta Q_{\alpha,\beta} & \geq & T_R \left(S_\alpha-S_\beta+ k\ln
 \func{P(\beta|\alpha)}\right)
 \end{eqnarray}

\subsubsection{Expectation values}
The problem with optimising for an individual transition is that
this can go catastrophically wrong if the operation is performed
upon any of the other $\alpha^\prime$ input logical states.  For
logically irreversible processes, as $w_{\alpha^\prime} \rightarrow
0$, then $\Delta W_{\alpha^\prime,\beta} \rightarrow \infty$.

We need to consider an optimisation over the full set of input
logical states, rather than with respect to a single input logical
state.  For the set of all possible transitions, we will seek to
minimise the expectation value, or mean cost, of performing the
operation.

This is not the only criteria that could be used.  One may seek
instead, for example, to optimise by a minimax criteria: minimising
the maximum cost that might be incurred.  This would lead to a
different set of $w_\alpha$ to those we will calculate here.  The
maximum cost that might be incurred with such a set would, for
certainty, be no higher than the maximum cost we will arrive at
here.  However, the expectation value for the cost, with the
different set, would be at least as high as the expectation value we
will find.

To be able to calculate an expectation value, a probability
distribution over the input logical states is needed.  For this we
will use the probabilities that go into the calculation of the
Shannon information of the input state: $P(\alpha)$.  The
probability of the transition $\alpha \rightarrow \beta$ occurring
is then $P(\beta|\alpha)P(\alpha)$ and the expectation values for
the work requirement is:
\begin{eqnarray}\label{eq:optwork}
 \mean{\Delta W} & = &\sum_\beta P(\beta) \left(E_\beta -T_R
 S_\beta\right)
 -\sum_\alpha P(\alpha) \left(E_\alpha -T_R S_\alpha\right) \nonumber \\
  & & +k T_R \sum_{\alpha,\beta} P(\alpha,\beta) \ln
 \func{\frac{w_\beta}{w_\alpha}}
 \end{eqnarray}
where $P(\alpha,\beta)=P(\beta|\alpha)P(\alpha)$ and
$P(\beta)=\sum_\alpha P(\alpha,\beta)$.

For logically reversible transformations, this is fixed:
\begin{equation}
\frac{w_\beta}{w_\alpha}=P(\beta|\alpha)=\frac{P(\beta)}{P(\alpha)}
\end{equation}

For logically irreversible transformations, we must vary the
$w_\alpha$ to minimise the function
\begin{equation}
X=\sum_{\alpha,\beta} P(\alpha,\beta) \ln
 \func{\frac{w_\beta}{w_\alpha}}
 \end{equation}
Consider the similar function
\begin{eqnarray}
Y&=&\sum_\beta P(\beta) \ln P(\beta)-\sum_\alpha P(\alpha) \ln P(\alpha) \nonumber  \\
 &=&\sum_{\alpha,\beta} P(\alpha,\beta) \ln \func{\frac{P(\beta)}{P(\alpha)}}\\
 X-Y&=&\sum_{\alpha,\beta} P(\alpha,\beta) \ln \func{\frac{w_\beta P(\alpha)}{P(\beta) w_\alpha}  \nonumber }\\
    &=&\sum_{\alpha,\beta} P(\alpha,\beta) \ln \func{\frac{P(\alpha,\beta)}{P(\beta) w(\alpha|\beta)} } \nonumber \\
    &\geq&0
 \end{eqnarray}
where
\begin{equation}
w(\alpha|\beta)=\frac{P(\beta|\alpha) w_\alpha}{w_\beta}
\end{equation}
and the equality occurs iff $P(\alpha,\beta) =P(\beta)
w(\alpha|\beta)$.  As $Y$ is independant of the values of the
$w_\alpha$, then the minimum value of $X$ is precisely the value of
$Y$. This minimum value of $X$ is reached when $w_\alpha=P(\alpha)$,
which leads to $w_\beta=P(\beta)$.

The result can easily be re-expressed as:

\begin{eqnarray}\label{eq:wprinciple}
 \mean{\Delta W} & \geq &\sum_\beta P(\beta) \left(E_\beta -T_R
 \left(S_\beta - k \ln P(\beta)\right)\right) \nonumber \\
 & &  -\sum_\alpha P(\alpha) \left(E_\alpha -T_R \left(S_\alpha - k \ln P(\alpha) \right) \right)
 \end{eqnarray}
This is the minimum expectation value of the work requirement for
the logical operation, using the physical procedure we have
described.  The same expression holds for logically reversible,
irreversible, deterministic and indeterministic operations.  It is
not hard to see that this also minimises the expectation value of
the heat generated:

\begin{eqnarray}\label{eq:principle}
 \mean{\Delta Q} & \geq & - T_R \left( \sum_\beta P(\beta) \left(S_\beta - k \ln
 P(\beta)\right) \right. \nonumber \\
 & & \left. -\sum_\alpha P(\alpha) \left(S_\alpha - k \ln P(\alpha) \right) \right)
\end{eqnarray}
As was noted for the case of $LE$ in \cite{Mar05b}, to achieve the
optimal physical implementation of a logically irreversible
operation requires the physical process to be designed for the
particular probability distribution $P(\alpha)$ over the input
logical states\footnote{It is worth noting that this is not the same
as having a prior knowledge of the input logical states.  Having
prior knowledge of which input state occurs allows one, trivially,
to do rather better than this, by choosing $w_\alpha^\prime=0$ for
all other input states.  This optimises for all individual
transitions that come from the known $\alpha$ input state, but
requires a different physical implementation each time a different
input logical state occurs.  That different physical implementation
is, in each case, equivalent to a logically reversible operation.}.
A physical implementation optimised for one input probability
distribution will not, in general, be optimised for a different
input probability distribution.  For logically irreversible
operations it is only possible to thermodynamically optimise the
logical transformation of information (where the input probability
is specified).  Without a probability distribution (even a default
assumption of equiprobable input states) it does not even make sense
to talk about optimising the expectation value for the work or heat
requirements, or about the Shannon information of the input and
output states.

\subsubsection{Multiple Heat Baths} For completeness, we note that if
there are several heat baths available, at different temperatures,
the equations may be easily generalised.  Defining:
\begin{eqnarray}
\overline{\mean{\Delta Q}}&=&\sum_i \mean{\Delta Q_i} \\
\overline{T}&=&\frac{\sum_i \mean{\Delta Q_i}
}{\sum_i\frac{\mean{\Delta Q_i }}{T_i}} \end{eqnarray} where
$\mean{\Delta Q_i}$ is the mean heat generated in a heat bath at
temperature $T_i$, we may simply replace $T_R$ with $\overline{T}$
and $\mean{\Delta Q}$ with $\overline{\mean{\Delta Q}}$, in Equation
\ref{eq:wprinciple}, and all subsequent equations.  In effect, this is equivalent to the possibility of using reversible Carnot cycles to rearrange heat between any heat baths available, in addition to performing the logical operation with a single heat bath.

The introduction of multiple heat baths has little practical significance though.  If
\begin{equation}
\sum_\beta P(\beta) \left(S_\beta - k \ln P(\beta)\right)
-\sum_\alpha P(\alpha) \left(S_\alpha - k \ln P(\alpha) \right) < 0
\end{equation}
then the least work is required by generating all the heat in the
coolest heat bath available.  If
\begin{equation}
\sum_\beta P(\beta) \left(S_\beta - k \ln P(\beta)\right)
-\sum_\alpha P(\alpha) \left(S_\alpha - k \ln P(\alpha) \right) > 0
\end{equation}
the opposite is true.  The least work involves only generating heat
in the hottest heat bath.

\subsection{Optimum physical process}\label{ss:nobetter}
We have shown that a particular physical process can implement a
logical operation, with a minimum expectation value for the work
required or heat generated.  Perhaps other physical processes might
exist which can perform the same logical operation at a lower cost?
We will now prove that {\em no} physical process can implement the
same logical transformation of information at a lower cost.

The initial statistical state of the logical processing apparatus is
\begin{equation}
\rho_I=\sum_\alpha P(\alpha) \rho_\alpha
\end{equation}
The final statistical state is
\begin{equation}
\rho_F=\sum_\beta P(\beta) \rho_\beta
\end{equation}
We assume that the environment is initially well described by a
canonical thermal state $\rho_E(T_R)$, at temperature $T_R$, and
that it is uncorrelated with the initial state of the logical
processing system.

Now consider the initial density matrix of the joint system of the
logical processing system and the apparatus
\begin{equation}
\rho=\rho_I \otimes \rho_E(T_R)
\end{equation}
so
\begin{equation}
\trace{\rho \ln \func{\rho}}=\trace{\rho_I \ln \func{\rho_I}}+
\trace{\rho_E(T_R) \ln \func{\rho_E(T_R)}}
\end{equation}

For any unitary evolution upon the combined system to be a physical
representation of the logical state, it must evolve the system to
some state $\rho^\prime$ such that the marginal distribution of the
information processing apparatus is:
\begin{equation}
\rho_F=\partrace{\rho^\prime}{E}
\end{equation}
The marginal distribution of the environment is then:
\begin{equation}
\rho_E^\prime=\partrace{\rho^\prime}{F}
\end{equation}

From the well known\cite{Tol1938,Weh78,Par89b} properties of unitary
evolutions and density matrices:
\begin{equation}
\trace{\rho \ln \func{\rho}}=\trace{\rho^\prime \ln
\func{\rho^\prime}}
\end{equation}

\begin{equation}
\trace{\rho^\prime \ln \func{\rho^\prime}} \geq \trace{\rho_F \ln
\func{\rho_F}}+ \trace{\rho_E^\prime \ln \func{\rho_E^\prime}}
\end{equation}
As $\rho_E(T_R)$ is a canonical distribution
\begin{equation}
\frac{H_E}{k T_R}=-\ln \func{\rho_E(T_R)} - \ln Z
\end{equation}
so

\begin{widetext}
\begin{equation}
\trace{\rho_E^\prime \left(\ln \func{\rho_E^\prime}+\frac{H_E}{k
T_R}\right)}- \trace{\rho_E(T_R)\left(\ln \func{\rho_E(T_R)}
+\frac{H_E}{k T_R}\right)} = \trace{\rho_E^\prime \left(\ln
\func{\rho_E^\prime}-\ln \func{\rho_E(T_R)}\right)} \geq 0
\end{equation}
\end{widetext} where $H_E$ is the internal Hamiltonian of the environment.  A
simple rearrangement gives
\begin{equation}
\trace{\rho_I \ln \func{\rho_I}}-\trace{\rho_F \ln \func{\rho_F}}
\geq \frac{\trace{H_E \rho_E(T_R)}}{k T_R}-\frac{\trace{H_E
\rho_E^\prime}}{k T_R}
\end{equation}
As the physical representations of the logical states are
non-overlapping:
\begin{eqnarray}
-k \trace{\rho_I \ln \func{\rho_I}}&=& \sum_\alpha P(\alpha)
\left(S_\alpha - k \ln P(\alpha) \right) \\
-k \trace{\rho_F \ln \func{\rho_F}} &=& \sum_\beta P(\beta)
\left(S_\beta - k \ln P(\beta)\right)
\end{eqnarray}

The expectation value for the work performed upon the system must
equal\footnote{We assume that the interaction energy between system
and environment is negligible at the start and end of the operation.
 Both this assumption, and the assumption that the environment is initially an uncorrelated Gibbs state, do
 not appear to hold in \cite{AN2001,AN2002b}.} the expectation value for the change in the internal energy of the
system plus the expectation value for the change in the internal
energy of the environment:
\begin{eqnarray}
\mean{\Delta W} &=& \sum_\beta P(\beta) E_\beta+\trace{H_E
\rho_E^\prime} \eqnwrap -\sum_\alpha P(\alpha)E_\alpha -\trace{H_E
\rho_E(T_R)}
\end{eqnarray}

From this we conclude that, for {\em any} physical process, which
takes input logical states $\{\alpha\}$ with probabilities
$P(\alpha)$ and produces output logical states $\{\beta\}$ with
probabilities $P(\beta)$, then the expectation value of the work
requirement for this process cannot be less than
\begin{eqnarray}
 \mean{\Delta W} & \geq & \sum_\beta P(\beta) \left(E_\beta -T_R
 \left(S_\beta - k \ln P(\beta)\right)\right) \eqnwrap
 -\sum_\alpha P(\alpha) \left(E_\alpha -T_R \left(S_\alpha - k \ln P(\alpha) \right) \right)
 \end{eqnarray}
There is no physical process that can do better, in terms of an
expectation value for the work requirement, or for the heat
generation, than the process developed in Section \ref{ss:genlogop}.

 We emphasise that the
relationships we have derived in this section do not depend upon the
results of the specific process we examined in the previous section.
No assumptions are made regarding the details of the physical
process which represents the logical operation, beyond the
requirements that it is a unitary evolution of the combined state
space of system and environment and does, in fact, faithfully
represent the operation. No assumptions are required about the
physical representation of the input and output logical states,
except those made in Section \ref{ss:physrepstates}. It is not
assumed that the environment is an ideal heat bath, is in some
thermodynamic limit, or is in thermal equilibrium after the
operation. The results require only that the environment be a
canonically distributed and uncorrelated system at the start of the
operation.  Given these assumptions, the result follows: there is no
physical representation of the logical operation that has a lower
expectation value for the work requirements or heat generation.

\section{Generalised Landauer's
principle}\label{ss:glpversions}

There are several different, but formally equivalent, ways of
expressing the Generalised Landauer's principle (GLP).  It will be
convenient to use the notation:

\begin{eqnarray}
\mean{\Delta E} &=& \sum_\beta P(\beta) E_\beta-\sum_\alpha
P(\alpha)E_\alpha \nonumber \\
\Delta S &=& \sum_\beta P(\beta) \left(S_\beta - k \ln
P(\beta)\right)
\eqnwrap -\sum_\alpha P(\alpha) \left(S_\alpha - k
\ln
P(\alpha) \right) \nonumber \\
\Delta H &=& -\sum_\beta P(\beta) \log P(\beta) +\sum_\alpha
P(\alpha) \log P(\alpha)
\end{eqnarray}
for: the change in the expectation value for of the internal energy
of the information processing apparatus; the change in the Gibbs-von
Neumann entropy of the statistical ensemble describing the
information processing system; and the change in the Shannon
information of the logical states over the course of the operation.

\subsection{Work requirements}

\noindent\textbf{GLP1: Work}
\begin{quote}
A logical transformation of information has a minimal expectation
value for the work requirement given by:
\begin{equation}
\mean{\Delta W} \geq \mean{\Delta E} - T_R \Delta S
\end{equation}
\end{quote}

\subsection{Heat generation}

Noting that \begin{equation} \mean{\Delta Q}=\mean{\Delta W} -
\mean{\Delta E} \end{equation} is equal to the expectation value of
the heat generated in the heat bath:

\noindent\textbf{GLP2: Heat}
\begin{quote}
A logical transformation of information has a minimal expectation
value for the heat generated in the environment of:
\begin{equation}\label{eq:glp2}
\mean{\Delta Q} \geq - T_R\Delta S
\end{equation}
\end{quote}

It is important to remember that the term $\Delta S$ appearing in
$GLP1$ and $GLP2$ is \textit{ not} the change in Shannon information
$\Delta H$ between the input and output states.  It is the change in
the Gibbs-von Neumann entropy of the logical system, taking into
account any changes in the entropies of the subensembles that
represent the input and output logical states.  It can be related to
the change in the Shannon information by
\begin{eqnarray}
\Delta S&=&\sum_\beta P(\beta)S_\beta - \sum_\alpha P(\alpha)
S_\alpha +k \Delta H \ln 2
\end{eqnarray}

\subsection{Entropic cost}
The change in the Gibbs-von Neumann entropy of the environmental
heat bath is given by:
\begin{equation}
\Delta S_{HB}=-k\trace{\rho_E^\prime \ln \func{\rho_E^\prime}}
+k\trace{\rho_E(T_R)\ln \func{\rho_E(T_R)}}
\end{equation}
which gives the entropic form of the Generalised Landauer's
principle:

\noindent\textbf{GLP3: Entropy}
\begin{quote}
A logical transformation of information requires a minimal change in
the Gibbs-von Neumann entropies of the marginal statistical states
of an information processing apparatus $\Delta S$ and its
environment $\Delta S_{HB}$ of:
\begin{equation}\label{eq:glp3}
\Delta S_{HB}+\Delta S \geq 0
\end{equation}
\end{quote}

This is a trivial consequence of the requirements that the evolution
be unitary and that the statistical states of the logical processing
system and the environment be initially uncorrelated.   The
expectation value of the heat generated in the environment is at
least equal to the increase in the Gibbs-von Neumann entropy of the
marginal state of the heat bath:
\begin{equation}
\mean{\Delta Q} \geq T_R \Delta S_{HB}
\end{equation}
This allows us to deduce $GLP1$ or $GLP2$ from $GLP3$, but not
reverse\footnote{In the limiting case of an {\em ideal} heat bath
and quasistatic processes, the equality is reached and the deduction
can then go in both directions.}.

\subsection{Information}

 If we
{\em define} the term:
\begin{eqnarray}\label{eq:nibdofapp}
\Delta S_{L}&=&\sum_\beta P(\beta)S_\beta - \sum_\alpha P(\alpha)
S_\alpha
\end{eqnarray}
we get
\begin{eqnarray}\label{eq:nibdof}
\Delta S_{HB}+\Delta S_{L} &\geq& -k \Delta H \ln 2
\end{eqnarray}
This expression seems suggestive.  If we regard the terms $\Delta
S_{HB}$ and $\Delta S_{L}$ as changes in the entropies of the
`non-information bearing degrees of freedom' of the environment and
the apparatus, respectively, then we appear to have provided a
quantitative version of Bennet's statement that

\begin{quote}
any logically irreversible manipulation of information $[\Delta H]$
\ldots must be accompanied by a corresponding $[k \ln 2]$ entropy
increase in the non-information bearing degrees of freedom of the
information processing apparatus $[\Delta  S_{L}]$ or its
environment $[\Delta S_{HB}]$ {\raggedleft \cite{Ben03}}
\end{quote}
although unlike Bennett, we do not restrict this to irreversible
transformations of data.

This produces what may be taken as the information form of the GLP:

\textbf{GLP4: Information}
\begin{quote}
A logical transformation of information requires an increase of
entropy of the non-information bearing degrees of freedom of the
information processing apparatus and its environment of at least $-k
\ln 2$ times the change in the total quantity of Shannon information
over the course of the operation:
\begin{eqnarray}\label{eq:glp4}
\Delta S_{NIBDF} &\geq& -k \Delta H \ln 2
\end{eqnarray}
\end{quote}

where $\Delta S_{NIBDF}=\Delta S_{HB}+\Delta S_{L}$.  This is quite
generally true and follows directly from $GLP3$ and the definition
of $\Delta S_L$.

\section{Models of computing}\label{s:glp}

We will now discuss some of the consequences that can be drawn from
the Generalised Landauer's principle by varying the thermodynamic
properties of the input and output states. This allows us to
consider the effects of having different energies and entropies for
the physical states that embody the logical states and has some
surprising consequences.



\subsection{Uniform Computing}\label{sss:uniform}
When we make the assumption that the computation takes place at the
same temperature throughout, such that

\begin{quote}
\refstepcounter{assumption} \label{as:isothermal}
\noindent\textbf{Assumption \arabic{assumption}:
Isothermal}
\begin{equation}\label{eq:assisothermal} \forall
\alpha,\beta \  T_R=T_\alpha=T_\beta
\end{equation}
\end{quote}
then we shall call this \textit{ isothermal} computing.

In the most commonly encountered set of assumptions for the
thermodynamics of computation, we have, in addition to the
assumption of isothermal computing, the physical states, that
represent the logical states, all have the same entropy and mean
energy, so that
\begin{quote}
\refstepcounter{assumption} \label{as:uniformstates}
\noindent\textbf{Assumption \arabic{assumption}: Uniform states}
\begin{eqnarray}
\forall \alpha,\beta \   E_R&=&E_\alpha=E_\beta \nonumber \\
\forall \alpha,\beta \  S_R&=&S_\alpha=S_\beta
\label{eq:uniformstates}
\end{eqnarray}
\end{quote}.

This reduces the Generalised Landauer's principle to the form of:
\begin{eqnarray}
\mean{\Delta W} &\geq& -k T_R \Delta H \ln 2 \nonumber \\
\mean{\Delta Q} &\geq& -k T_R \Delta H \ln 2 \label{eq:uniformglp}
\end{eqnarray}
where $\Delta H$ is the change in Shannon information over the
course of the transformation.  This is the usual form in which
Landauer's principle is encountered.

The necessary and sufficient conditions for these to hold is the
weaker condition:
\begin{quote}
\refstepcounter{assumption}\label{as:uniformcomp}
\noindent\textbf{Assumption \arabic{assumption}: Uniform computing}
\begin{eqnarray}
\sum_\alpha P(\alpha) E_\alpha &=&\sum_\beta
P(\beta) E_\beta \nonumber \\
\sum_\alpha P(\alpha) S_\alpha&=&\sum_\beta P(\beta) S_\beta
\label{eq:uniformcomp}
\end{eqnarray}
\end{quote}

\subsection{Equilibrium Computing} The simplifying assumption of
uniform computing is made so universally, that it might be
questioned whether there is any value to considering non-uniform
computing.  To answer this, consider what happens if the input and
output states are constructed to be canonical thermal systems, at
temperature $T_R$, with the properties:
\begin{quote}
\refstepcounter{assumption}\label{as:equilibrium} \textbf{Assumption
\arabic{assumption}: Equilibrium Computing}
\begin{eqnarray}\label{eq:assequilibrium}
E_\alpha-T_R S_\alpha +k T_R \ln P(\alpha) &=& C_A \nonumber \\
E_\beta-T_R S_\beta +k T_R \ln P(\beta)&=& C_A
\end{eqnarray}
\end{quote}
where $C_A$ is a constant, related to the overall size of the
logical processing apparatus.  This yields the relationships
\begin{equation}
\mean{\Delta E}-T_R\Delta S=0
\end{equation}
and reduces the Generalised Landauer's principle to

\begin{equation}\label{eq:nowork}
\mean{\Delta W} \geq 0
\end{equation}
although
\begin{equation}
\mean{\Delta Q} \geq -T_R\Delta S
\end{equation}
still.  The equality can, of course, only be reached in the limit of
slow processes.

The necessary and sufficient assumption for Equation \ref{eq:nowork}
to hold is
\begin{quote}
\refstepcounter{assumption}\label{as:nomeanwork} \textbf{Assumption
\arabic{assumption}: Zero mean work}
\begin{eqnarray}\label{eq:nomeanwork}
&\sum_\alpha P(\alpha) \left(E_\alpha-T_R \left(S_\alpha +k \ln
P(\alpha) \right)\right)& \nonumber \\
& = \sum_\beta P(\beta) \left(E_\beta-T_R \left(S_\beta +k \ln
P(\beta)\right)\right)&
\end{eqnarray}
\end{quote}
Assumption \ref{as:nomeanwork} only implies the average work
requirement can approach zero, over all the possible transitions
between logical states.  Assumption \ref{as:equilibrium} ensures
that there is a zero mean work requirement $\Delta W_{\alpha,\beta}$
for all individual $(\alpha,\beta)$ transitions.

\subsection{Adiabatic Computing}
To eliminate mean heat generation in the ideal limit, the necessary
and sufficient condition is:
\begin{quote}
\refstepcounter{assumption}\label{as:nomeanheat} \textbf{Assumption
\arabic{assumption}: Zero mean heat generation}
\begin{equation}\label{eq:nomeanheat}
\sum_\alpha P(\alpha) \left(S_\alpha +k \ln P(\alpha) \right)
 = \sum_\beta P(\beta) \left(S_\beta +k \ln P(\beta)\right)
\end{equation}
\end{quote}
leading to
\begin{equation}\label{eq:noheat}
\mean{\Delta Q} \geq 0
\end{equation}
although this does not eliminate mean work requirements
\begin{equation}
\mean{\Delta W} = \mean{\Delta E}
\end{equation}
Again, this is only the expectation value over all transitions.  To
ensure that the mean heat generated $\Delta Q_{\alpha,\beta}$ is
zero for each individual $(\alpha,\beta)$ transition, requires:
\begin{quote}
\refstepcounter{assumption}\label{as:adiabatic} \textbf{Assumption
\arabic{assumption}: Adiabatic computing}
\begin{eqnarray}
S_\alpha +k \ln P(\alpha) &=& C_B \nonumber \\
S_\beta +k \ln P(\beta)&=& C_B \label{eq:adiabatic}
\end{eqnarray}
\end{quote}
where $C_B$ is an apparatus related constant.

\subsection{Adiabatic Equilibrium Computing}
Combining the assumptions of adiabatic and equilibrium computing
gives the requirement
\begin{quote}
\refstepcounter{assumption}\label{as:adiabaticequilibrium}
\textbf{Assumption \arabic{assumption}: Adiabatic equilibrium
computing}
\begin{eqnarray}\label{eq:adiabaticequilibrium}
E_\alpha=E_\beta &=&E_R \nonumber \\
S_\alpha +k \ln P(\alpha) &=& C_C \nonumber \\
S_\beta +k \ln P(\beta)&=& C_C
\end{eqnarray}
\end{quote}
which yields, $\forall \alpha,\beta$
\begin{eqnarray}
\Delta W_{\alpha,\beta}& \geq &0 \nonumber \\
\Delta Q_{\alpha,\beta}& \geq &0
\end{eqnarray}
with equality being reachable as a limiting case, and $C_C$ again a
machine dependant constant.

 This result may
seem surprising. It suggests that it is possible to design a
computer to perform any combination of logical operations, with no
exchange of heat with the environment and requires no work to be
performed upon it.  This must be as true for logically irreversible
operations as for logically reversible operations, and as true for
logically indeterministic operations as for logically deterministic
operations.

To understand this better, let us consider what happens in adiabatic
equilibrium computing.  We can use the square well potential as the
physical model of the logical states, as the internal energy of
these states is $\frac{1}{2}kT$.  Varying the width of the square
well potential for each input and output logical state satisfies the
remaining conditions.

Implementing the model of adiabatic equilibrium computing on the
processes of Section \ref{s:theproof} simplifies the procedure
significantly:
\begin{enumerate}
\item There is no need to resize the input states, as these will already be canonically distributed.
Steps 1 through to 3 are redundant.
\item Potential barriers are inserted into the $\alpha$ states, corresponding to the
conditional probabilities $P(\beta|\alpha)$, as in Step 4.
\item The separate portions of the $\beta$ output states are brought into adjacent positions as in
Step 5.
\item The potential barriers within each $\beta$ output states are removed, as in Step
6.
\item These output states are \textit{ already} canonically distributed.  There is therefore no need for
a resizing of the output states and Step 7 through to 9 are
unnecessary.
\end{enumerate}
None of these stages require any work to be performed upon the
system or exchange of heat with the environment.  The computation is
reduced to a process of rearranging a canonical ensemble from one
set of canonically distributed orthogonal subensembles into a
different set of canonically distributed orthogonal subensembles, in
accordance with the computational probabilities $P(\beta|\alpha)$.

As the probabilities of the different output states cannot change
between logical operations\footnote{By definition anything that
changes the probabilities of a state must be a logical
transformation of the data.} then the canonically distributed output
states can be used as canonically distributed input states to any
new logical operation.  This thermodynamic model may therefore
proceed indefinitely without generating any heat or requiring any
work.

Before leaving this subject, let us just note one feature of
equilibrium computing.  Logically deterministic, irreversible
computations are able to avoid generating heat, in this model, by
increasing the size of the physical states representing the logical
states.  This does \textit{ not} mean that the logical processing
apparatus itself needs to be increasing in size.  Although the size
of the individual states has increased, the number of logical states
has decreased (by the definition of a logically deterministic,
irreversible computation!).  Whenever the equality in Assumption
\ref{as:nomeanheat} holds, the two effects cancel out and the
overall size of the logical processing apparatus can remain
constant.
\section{Thermodynamic reversibility}
We have not yet examined the question of whether these operations
are thermodynamically reversible.  This is a subtle question and
depends upon what one takes to be the statistical mechanical
generalisation of thermodynamic entropy and thermodynamic
reversibility.  We will first discuss how this appears from the
perspective of three different approaches to entropy, and then from
a definition based on thermodynamic cycles that is not directly
based upon any definition of entropy.

It is worth remembering that a net increase in entropy is considered
is taken as a sign of \textit{irreversibility} because net decreases
in entropy cannot occur (or are unlikely).  An `entropy' that can be
\textit{systematically} decreased may be a useful indicator of some
properties, but its increase cannot automatically be regarded as an
indicator of irreversibility, whether thermodynamic or of some other
kind.

We will consider three possible conditions for thermodynamic
reversibility and irreversibility:
\begin{enumerate}
\item The thermodynamic entropy is the entropy of the individual
state.  If the system is in logical state $\alpha$, then the
thermodynamic entropy is $S_\alpha$.  The net entropy change for a
particular logical transition, from logical state $\alpha$ to
logical state $\beta$ is:
\begin{equation}\label{eq:individualentropy}
S_\beta-S_\alpha+\frac{\Delta Q_{\alpha,\beta}}{T_R}
\end{equation}

A transition is thermodynamically reversible if the decrease in
individual state entropy from the input to output logical states is
equal to the heat generated in the heat bath, divided by the
temperature of the heat bat.  A transition is thermodynamically
irreversible if the decrease in individual state entropy is less
than this. Decreases in individual state entropy greater than this
cannot occur.
\item The thermodynamic entropy is the entropy of the individual
state, but is only non-decreasing \textit{on average}.  If the
system is in logical state $\alpha$, then the thermodynamic entropy
is $S_\alpha$, but this may decrease provided it does not decrease
on average.  The average change is:
\begin{equation}
\sum_\beta  P(\beta) S_\beta-\sum_\alpha P(\alpha)
S_\alpha+\frac{\mean{\Delta Q_{\alpha,\beta}}}{T_R}
\end{equation}

A logical transformation of information is thermodynamically
reversible if the average decrease in individual state entropy over
all the transitions from input to output logical states is equal to
the average heat generated in the heat bath, divided by the
temperature of the heat bath.  The transformation is
thermodynamically irreversible if the average decrease in individual
state entropy is less than this. Average decreases in individual
state entropy greater than this cannot occur.
\item The thermodynamic entropy is the Gibbs-von Neumann entropy of the \textit{ marginal} statistical states.
  If the statistical state of the system is $\rho=\sum_\alpha P(\alpha)\rho_\alpha$, the
  thermodynamic entropy is $-k \trace{\rho \ln \func{\rho}}$.  A logical
transformation of information is thermodynamically reversible if the
decrease in Gibbs-von Neumann entropy from the input to output
statistical states is equal to the average heat generated in the
heat bath, divided by the temperature of the heat bath.  The
transformation is thermodynamically irreversible if the decrease in
Gibbs-von Neumann entropy is less than this.  Decreases in Gibbs-von
Neumann entropy greater than this cannot occur.
\end{enumerate}
The first two conditions imply thermodynamic irreversibility for
logically deterministic, irreversible operations.  Unfortunately, it
will be shown neither condition can consistently account for
logically indeterministic operations, which can systematically
decrease the relevant entropy measure by quantities greater than
should be permitted.

The third condition gives an entropy that is consistently
non-decreasing (provided there are no spontaneous or pre-existing
correlations with heat baths).  Logically indeterministic operations
do not decrease this entropy.  On the other hand, logically
irreversible operations no longer necessarily increase this entropy
measure either.  According to the Gibbs-von Neumann measure, all
logical operations may be implemented in a thermodynamically
reversible manner.

We will be making the standard assumptions that all processes can
take place with ideal heat baths and sufficiently slowly that
equalities are reached as the limiting cases.  Without these
assumptions no process can be thermodynamically reversible.  We will
therefore replace the appropriate inequalities with equalities.

\subsection{Individual logical state entropy}
The net individual state entropy change, for a particular logical
transition, gives:
\begin{equation}
S_\beta-S_\alpha+\frac{\Delta Q_{\alpha,\beta}}{T_R} \geq k\ln
 \func{P(\beta|\alpha)}
\end{equation}
Allowed logically deterministic transitions require
$P(\beta|\alpha)=1$. The equality is automatically reached for
logically deterministic, reversible transitions, and which are
therefore thermodynamically reversible. For logically deterministic,
irreversible transitions, the equality requires $w_\alpha=1$. This
is only possible if no other input logical states are allowed. Such
an operation would be trivially logically reversible as there is
only one permissible input logical state. So according to this
entropy measure, logically deterministic irreversible transitions
must be thermodynamically irreversible.

As $\ln
 \func{P(\beta|\alpha)} \leq 0$ it is possible that
\begin{equation}
S_\beta-S_\alpha+\frac{\Delta Q_{\alpha,\beta}}{T_R} < 0
\end{equation}
This gives a net decrease in individual state entropy.  For this to
happen, the transition must be logically indeterministic. Optimally
implemented, logically indeterministic, reversible transitions will
\textit{ always} decrease individual state entropy.

If an entropy increase is indicative of thermodynamic
irreversibility because entropy decreases are impossible, this
measure of entropy cannot be seen as a good indicator of
thermodynamic irreversibility.  Any apparent irreversibility can
actually be reversed.

\subsection{Average state entropy} In
statistical mechanics fluctuations occur.  Perhaps the demand for a
\textit{strictly} non-decreasing entropy might be the problem.  What
of the \textit{average} change in entropy?  Does this give a good
indicator of thermodynamic irreversibility?

This gives:
\begin{equation}
\sum_\beta  P(\beta) S_\beta-\sum_\alpha P(\alpha)
S_\alpha+\frac{\mean{\Delta Q_{\alpha,\beta}}}{T_R}=-k \Delta H \ln
2
\end{equation}
What we have here is the ideal limit case of $GLP4$, with $\Delta
S_{NIBDF}$ now representing the average change in entropy:
\begin{equation}
\Delta S_{NIBDF}=\sum_\beta  P(\beta) S_\beta -\sum_\alpha P(\alpha)
S_\alpha +\frac{\mean{\Delta Q_{\alpha,\beta}}}{T_R}
\end{equation}

Logically deterministic, irreversible operations have $\Delta H < 0$
then
\begin{equation}
\Delta S_{NIBDF} > 0
\end{equation}
and the net mean change in individual state entropy of the system
and environment is strictly increasing.  Again logically
deterministic, irreversible operations must be, on average,
individual state entropy increasing.

The problem with the argument should be immediately apparent: for
logically reversible, indeterministic operations $\Delta H > 0$ and
by the same reasoning and arguments it is possible that
\begin{equation}
\Delta S_{NIBDF} = -k \Delta H \ln 2  < 0
\end{equation}
Not only can logically indeterministic operations reduce individual
state entropy on individual transitions, they can even reduce this
entropy {\em on average}.

\subsection{Gibbs-von Neumann entropy}\label{ss:gvnentropy}
The entropy measure which includes the effects of the statistical
mixture over the states, the Gibbs-von Neumann entropy over the
ensemble, gives the initial entropy of the logical processing
system:
\begin{equation}
S_I=-k \trace{\rho_I \ln \func{\rho_I}}
\end{equation}
where
\begin{equation}
\rho_I=\sum_\alpha P(\alpha) \rho_\alpha
\end{equation}
and the final entropy:
\begin{equation}
S_F=-k \trace{\rho_F \ln \func{\rho_F}}
\end{equation}
where
\begin{equation}
\rho_F=\sum_\beta P(\beta) \rho_\beta
\end{equation}

In \cite{Maroney2007a} it is argued that the Gibbs-von Neumann
entropy is indeed the correct statistical mechanical generalisation
of thermodynamic entropy, although this identification has not been
assumed anywhere within this paper\footnote{We have calculated the
Gibbs-von Neumann entropies for {\em individual} states, in
canonical distributions, but even here the calculation of the mean
work requirements and mean heat generated did not depend upon any
identification of this as a \textit{thermodynamic} entropy.}.

When we consider the Gibbs-von Neumann entropy, the most appropriate
form of the GLP is $GLP3$.  In this case, the limiting behaviour
gives
\begin{equation}
\Delta S_{HB}+\Delta S = 0
\end{equation}
As any logical operation may reach this limit, the Gibbs-von Neumann
entropy regards all logical operations as being possible in a
thermodynamically reversible manner.

\subsection{Discussion}
As some of these results may seem surprising or counter-intuitive,
and appear to contradict widely stated expressions of the
implications of the thermodynamics of logically irreversible
operations, let us examine them in more detail.

\subsubsection{The $RLE$-$LE$ cycle} First, let us take the examples of the logically
deterministic, irreversible Reset to Zero ($RTZ$) operation and the
logically reversible, indeterministic Unset from Zero ($UFZ$)
operation (Appendix \ref{ap:onebit}). If the argument is accepted
that the optimal procedure to implement $RTZ$ is entropy increasing,
then it must also be accepted that the optimal procedure for $UFZ$
can be entropy \textit{decreasing}.

That this must be the case can be seen by considering the Reverse
Landauer Erasure ($RLE$) operation immediately followed by the
Landauer Erasure ($LE$) operation.  If these two procedures are
matched in terms of the probabilities, input and output states, then
the result is to leave both the logical system and the environment
in their initial states.  The total entropy must be the same at the
end of such a procedure, as at the start, and it follows if it
\textit{increases} during $LE$, then it \textit{must} decrease
during $RLE$.

As a simple example, using the assumptions of uniform computing, and
an initial input state of $0$, the process of $RLE$ extracts $kT \ln
2 $ heat from the environment, and converts it into work.  The
output state of $RLE$ is an equiprobable distribution of logical
states $0$ and $1$, each of which has the same entropy as the
initial $0$ state.

This is input to the $LE$ procedure, which requires $kT \ln 2$ heat
to be generated in the environment and leaves the output state as
$0$.  The system and environment are left in the same logical and
thermodynamic states as at the beginning of the process. There is a
zero net work requirement and a zero net heat generation.  The
combination of $RLE$ followed by $LE$ is clearly a thermodynamically
reversible cycle.

It follows that the net change in entropy over the course of the two
operations must be zero for both system and environment. To argue
that the net change in entropy for the $LE$ procedure is $k \ln 2$,
requires, for the overall change in entropy to be zero, the change
in entropy during the $RLE$ operation to be $-k \ln 2$.

Both the individual state entropies, and the average state entropy,
do indeed decrease by $k \ln 2$ during the $RLE$ operation.  The
Gibbs-von Neumann entropy remains constant, as the mixing entropy
increases by $k \ln 2$ to compensate.  During the course of the $LE$
operation, the individual state, and average state, entropies
increase by $k \ln 2$.  In the conventional operation of the $LE$
process, this is associated with heat generated in the environment,
and is often considered to be the source of an irreversible entropy
increase.  However, we can clearly see that from the point of view
of the Gibbs-von Neumann entropy, there is a compensating reduction
of $k \ln 2$ associated with the reduction in the mixing entropy.

\subsubsection{Uniform Computing}
We can easily generalise this to situations where the quantity of
information erased is less that 1 bit\footnote{This cycle was
detailed in \cite{Mar05b}.}, and in doing so will see more clearly
the need to optimise the operation to the probability distribution.
We simply need to implement an $UFZ(p)$ operation, followed by an
$RTZ(p)$ operation.

We start with a standard ``atom in a box'', and the partition
divides the box exactly in half.  The atom is on the left hand side
(which represents logical state $0$) with certainty.
\begin{enumerate}
\item $RLE(p)$.

The $RLE(p)$ operation consists of the following steps:
\begin{enumerate}
\item Isothermally move the partition to the right hand side,
extracting $W_1=k T \ln 2$ heat as work.
\item Insert the partition at location $x=pL$ in the box, where the
width of the box is $L$.  The atom is, with probability $p$, on the
left hand side of the partition.
\item Isothermally move the partition to the centre of the box ($x= \half L$).  If
the atom is on the left hand side, the work requirement is $k T \ln
(2p)$ while if it is on the right, the work requirement is $k T \ln
(2(1-p))$.  The mean work required in this stage is
\begin{equation}
W_2=kT \left( p \ln p + (1-p) \ln (1-p) +\ln 2\right)
\end{equation}
so the net work for the operation is
\begin{equation}
W_1+W_2=kT \left( p \ln p + (1-p) \ln (1-p)\right)
\end{equation}
which is negative, representing a net extraction of work.
\end{enumerate}
Now, we find that the individual state entropy, and average state
entropy, remains the same as at the start of the operation, despite
the fact that $kT \left( p \ln p + (1-p) \ln (1-p)\right) $ work has
been extracted from the heat bath.  From the point of view of the
Gibbs-von Neumann entropy, this is compensated by the increase in
mixing entropy between the two logical states.

\item $LE(p)$.

If we follow this with an $LE(p)$, we have the steps:
\begin{enumerate}
\item Isothermally move the partition to the position $x=pL$.  If
the atom is on the left hand side, the work requirement is $-k T \ln
(2p)$ while if it is on the right, the work requirement is $-k T \ln
(2(1-p))$.  The mean work required in this stage is
\begin{equation}
W_3=-kT \left( p \ln p + (1-p) \ln (1-p) +\ln 2\right)
\end{equation}
\item Remove the partition from the box.
\item Insert the partition in the right hand side of the box and
isothermally move it to the centre.  This requires $W_4=k T \ln 2$
work, so the net work is
\begin{equation}W_3+W_4=-kT \left( p \ln p
+ (1-p) \ln (1-p)\right)
\end{equation}
\end{enumerate}
Again, both the individual and average state entropy are unchanged,
while work is converted to heat in the environment.  The Gibbs-von
Neumann entropy, however, shows a compensating decrease in mixing
entropy.

The net work and net heat generated, over the course of the cycle,
is zero:
\begin{equation}
W_1+W_2+W_3+W_4=0
\end{equation}
If the heat generated in the environment during the $LE(p)$
operation is an indicator of an irreversible entropy increase, we
have to explain a corresponding systematic \textit{reduction} in
entropy during the $RLE(p)$ operation.  As we noted, entropy
increases are associated with irreversibility precisely because
corresponding systematic entropy decreases are supposed to be
impossible.

\item $LE(p^\prime)$.

Let us now consider following the $RLE(p)$ operation with
$LE(p^\prime)$, where the erasure operation has been optimised for a
different probability distribution.
\begin{enumerate}
\item Isothermally move the partition to the position $x=p^\prime L$.  If
the atom is on the left hand side, the work requirement is $-k T \ln
(2p^\prime)$ while if it is on the right, the work requirement is
$-k T \ln (2(1-p^\prime))$.  The mean work required in this stage is
\begin{equation}
W_5=-kT \left( p \ln p^\prime + (1-p) \ln (1-p^\prime) +\ln 2\right)
\end{equation}
\item Remove the partition from the box.
\item Insert the partition in the right hand side of the box and
isothermally move it to the centre.  This requires $W_6=k T \ln 2$
work, so the net work is
\begin{equation}
W_5+W_6=-kT \left( p \ln p^\prime + (1-p) \ln (1-p^\prime) \right)
\end{equation}
\end{enumerate}
The net work required over the $RLE(p)$-$LE(p^\prime)$ cycle is
\begin{equation}
\begin{array}{c}W_1+W_2\\+W_5+W_6 \end{array}=kT \left( p \ln \left[ \frac {p}{ p^\prime}
\right]
 + (1-p) \ln \left[ \frac {1-p}{1-p^\prime} \right] \right) \geq 0
\end{equation}
with equality occurring if, and only if, $p=p^\prime$.

Once again, both the individual and average state entropy are
unchanged.  In this case, however, the cycle generates a net heat in
the environment, unless $p=p^\prime$.  This cycle is, in general,
thermodynamically irreversible.
\end{enumerate}

From the point of view of the Gibbs-von Neumann entropy, it is the
removal of the partition from the location $x=p^\prime L$, when the
probability is $p$, that associated with an uncompensated entropy
increase.  We can see this by noting that if we reinsert the
partition at $x=p^\prime L$, we do not recover the previous
statistical state, as the probability of the atom being on the left
hand side would then be $p^\prime$.  To recover the statistical
state we need to reinsert the partition at $x=p L$, then move it
isothermally to $x=p^\prime L$.  This isothermal movement of the
partition requires, on average:
\begin{equation*}
kT \left( p \ln \left[ \frac {p}{ p^\prime} \right]
 + (1-p) \ln \left[ \frac {1-p}{1-p^\prime} \right] \right) \geq 0
\end{equation*}
work to be performed.

Even so, let us note that had $LE(p^\prime)$ in fact followed the
$RLE(p^\prime)$ operation, it would have been thermodynamically
reversible.  The physical process involved in performing the $LE(p)$
(or $LE(p^\prime)$) operation cannot be said to be intrinsically
thermodynamically reversible (or irreversible) in itself.  Whether
it is thermodynamically reversible, or not, depends upon the
statistical state upon which it acts.

\subsubsection{Adiabatic Equilibrium Computing}
Let us look at the same logical cycle, but with a different
computing model: adiabatic equilibrium.  Again we start with a
standard ``atom in a box''. As the atom is in logical state $0$ with
certainty, the conditions of Equation \ref{eq:adiabaticequilibrium}
require that logical state $0$ occupies the entire box.
\begin{enumerate}
\item $RLE(p)$.

The $RLE(p)$ operation now consists of the single step:
\begin{enumerate}
\item Insert the partition at location $x=pL$ in the box, where the
width of the box is $L$.  The atom is, with probability $p$, on the
left hand side of the partition.
\end{enumerate}
No work is required or heat generated.  The individual and average
state entropies have decreased, with the average state entropy
decreasing by $k \left( p \ln p + (1-p) \ln (1-p)\right) $. The
Gibbs-von Neumann entropy remains the same, as the mixing entropy
compensates for this.
\item $LE(p)$.

If we follow this with an $LE(p)$, we have the step:
\begin{enumerate}
\item Remove the partition from the box.
\end{enumerate}
Both the individual and average state entropy are increased, with
the average state entropy increasing by $k \left( p \ln p + (1-p)
\ln (1-p)\right)$.  The Gibbs-von Neumann entropy, however, shows a
compensating decrease in mixing entropy.

We see how, in the case of adiabatic equilibrium computing, the
generation of heat in the environment is replaced by changes in the
entropies of the individual states (or, as \cite{Ben03} refers to
it, the non-information bearing degrees of freedom of the
apparatus).  Although there is an increase in such entropies during
the $LE(p)$ process, there is an exactly equivalent
\textit{decrease} during the $RLE(p)$ process.  Again, if we take
the increase during $LE(p)$ to be indicative of a thermodynamic
irreversibility, we are left with the challenge of accounting for
the systematic decrease during the $RLE(p)$ operation.

\item $LE(p^\prime)$

Following $RLE(p)$ with an $LE(p^\prime)$ operation under the
assumptions of adiabatic equilibrium does not entirely make sense,
as adiabatic equilibrium requires the physical representation of the
logical states is tailored to the probability of the state
occurring.  However, we may consider the optimum implementation of
$RTZ(p^\prime)$, on the assumption that the probability of the
logical state $0$ is $p^\prime$, with the partition initially
located at $x=p L$ and the process leaving the system in a state
compatible with adiabatic equilibrium computation.
\begin{enumerate}
\item Isothermally move the partition to the position $x=p^\prime L$.  If
the atom is on the left hand side, the work requirement is $-k T \ln
\left( \frac{p^\prime}{p} \right)$ while if it is on the right, the
work requirement is $-k T \ln \left(\frac{1-p^\prime}{1-p} \right)$.
The mean work required\footnote{Note that, had the probability of
logical state $0$ \textit{actually} been $p^\prime$, the work
required would have been:
\begin{equation}
kT \left( p^\prime \ln \left[ \frac {p}{ p^\prime} \right]
 + (1-p^\prime) \ln \left[ \frac {1-p}{1-p^\prime} \right] \right) \leq 0
\end{equation}
so work would have been extracted in the process.}
 in this stage is
\begin{equation}
kT \left( p \ln \left[ \frac {p}{ p^\prime} \right]
 + (1-p) \ln \left[ \frac {1-p}{1-p^\prime} \right] \right) \geq 0
\end{equation}
with equality occurring if, and only if, $p=p^\prime$.

\item Remove the partition from the box.
\end{enumerate}
The net work required over the $RLE(p)$-$LE(p^\prime)$ cycle is
again
\begin{equation}
kT \left( p \ln \left[ \frac {p}{ p^\prime} \right]
 + (1-p) \ln \left[ \frac {1-p}{1-p^\prime} \right] \right) \geq 0
\end{equation}
\end{enumerate}
Once again, from the point of view of the Gibbs-von Neumann entropy,
it is the removal of the partition from the location $x=p^\prime L$,
when the probability is $p$, that is associated with an
uncompensated entropy increase.

\subsubsection{Generic logical operations}
Now let us consider a generic logical transformation of information.
Start with input logical states ${\alpha}$, physically represented
by states with energies and entropies $E_\alpha$ and $S_\alpha$, and
define a logical operation by the transition probabilities
$P(\beta|\alpha)$ to the output logical states ${\beta}$ with
physical state energies and entropies $E_\beta$ and $S_\beta$.

To thermodynamically optimise the physical process, we need a
probability distribution $P(\alpha)$.  The $\beta$ output states
will then occur with probabilities
\begin{equation}
P(\beta)=\sum_\alpha P(\beta|\alpha) P(\alpha)
\end{equation}
Writing
\begin{eqnarray*}
\rho_I &=&\sum_\alpha P(\alpha) \rho_\alpha \\
\rho_F &=&\sum_\beta P(\beta) \rho_\beta
\end{eqnarray*}
then the optimal thermodynamic cost of this is:
\begin{eqnarray}
\mean{\Delta W}&=&\left(\trace{H\rho_I}-T_R S\func{\rho_I}\right) -
\left(\trace{H\rho_F}-T_R S\func{\rho_F}\right) \nonumber \\
 &=&\sum_\beta P(\beta) \left(E_\beta -T_R
 S_\beta\right)  -\sum_\alpha P(\alpha) \left(E_\alpha -T_R
 S_\alpha\right) \eqnwrap
 +k T_R \sum_{\alpha,\beta} P(\alpha,\beta) \ln
 \func{\frac{P(\beta)}{P(\alpha)}}
\nonumber \\
\mean{\Delta Q}&=&-T_R S\func{\rho_I}+T_R S\func{\rho_F}
\end{eqnarray}

We can now define a physical process, that acts upon the physical
states $\{\beta\}$, and evolves them into the physical states
$\{\alpha\}$, with probabilities\footnote{For logical operations
taking as input states $\{\beta\}$ and producing output states
$\{\alpha\}$, we will use the notation $\Pi$ for the corresponding
probabilities.} given by
\begin{equation}
\Pi(\alpha|\beta)=\frac{P(\beta|\alpha)P(\alpha)}{\sum_\alpha
P(\beta|\alpha)P(\alpha)}
\end{equation}
It is straightforward to see that if this acts upon states
$\{\beta\}$, occurring with probabilities $P(\beta)$, then it
produces the states $\{\alpha\}$ with probabilities $P(\alpha)$.  If
the physical process is optimised for these probabilities, then the
thermodynamic cost is
\begin{eqnarray}
\mean{\Delta W_{\Pi}}&=&-\mean{\Delta W}\nonumber \\
\mean{\Delta Q_{\Pi}}&=&-\mean{\Delta Q}
\end{eqnarray}
So for any logical transformation of information, optimally
implemented, there exist a second operation, which when optimally
implemented restores the original statistical state, and for which
the total expectation value of the work requirement, and the total
expectation value of the energy generated in the environment, is
zero.  This is true regardless of whether the original operation is
logical reversible, irreversible, deterministic or indeterministic.

As we have noted before, however, to achieve this optimum for
logically irreversible operations, the physical process must take
into account the probability distribution $P(\alpha)$ over the input
logical states.  One cannot create a physical process, that
implements a logically irreversible operation, which will be
thermodynamically optimal for every probability distribution over
the input logical states.  This differs from logically reversible
operations, which may be represented by a physical process which is
thermodynamically optimal for any probability distribution over the
input logical states.

We will now look at the effect of an operation that is not optimised
for the right set of probabilities.  Suppose we have an operation
with the same transition probabilities $\Pi(\alpha|\beta)$, above,
but the physical process has been optimised for the input
probability distribution $\Pi(\beta)$.  The output states are
expected to occur with probabilities
\begin{equation}
\Pi(\alpha)=\sum_\beta \Pi(\alpha|\beta) \Pi(\beta)
\end{equation}
and the expected thermodynamic cost, from Equation \ref{eq:optwork},
is:
\begin{eqnarray}
 \mean{\Delta W_\Pi} & = &\sum_\alpha \Pi(\alpha) \left(E_\alpha -T_R S_\alpha\right)
-\sum_\beta \Pi(\beta) \left(E_\beta -T_R
 S_\beta\right) \eqnwrap
  +k T_R \sum_{\alpha,\beta} \Pi(\alpha,\beta) \ln
 \func{\frac{w_\alpha}{w_\beta}}
 \end{eqnarray}
with $w_\alpha=\Pi(\alpha)$ $w_\beta=\Pi(\beta)$ and
$\Pi(\alpha,\beta)=\Pi(\alpha|\beta)\Pi(\beta)$.

The input states do not occur with $\Pi(\beta)$ but with $P(\beta)$.
The actual thermodynamic cost incurred is:
\begin{eqnarray}
 \mean{\Delta W^{\prime}_\Pi} & =&
 \sum_\alpha P(\alpha) \left(E_\alpha -T_R S_\alpha\right) -\sum_\beta P(\beta) \left(E_\beta -T_R
 S_\beta\right)\eqnwrap
 +k T_R \sum_{\alpha,\beta} P(\alpha,\beta) \ln
 \func{\frac{w_\alpha}{w_\beta}}
 \end{eqnarray}

The combined cycle now has a cost
\begin{equation}
 \mean{\Delta W^{\prime}_\Pi}+ \mean{\Delta W}=k T_R \sum_{\alpha,\beta} P(\alpha,\beta) \ln
 \func{\frac{w_\alpha P(\beta)}{w_\beta P(\alpha)}}
\end{equation}
which can be rearranged to give
\begin{equation}
 \mean{\Delta W^{\prime}_\Pi}+ \mean{\Delta W}=k T_R \sum_{\alpha,\beta} P(\alpha,\beta) \ln
 \func{\frac{P(\alpha,\beta)}{\Pi(\beta|\alpha)P(\alpha)}} \geq
 0
\end{equation}
where $\Pi(\beta|\alpha)\Pi(\alpha)=\Pi(\alpha|\beta)\Pi(\beta)$.

Equality can occur in two ways.  Firstly, and most simply, if
$\Pi(\beta)=P(\beta)$.  The input states to the $\Pi(\alpha|\beta)$
operation occur with the optimal probabilities.

Secondly, if the second operation is a logically reversible
operation, then
\begin{equation}
\forall \beta \; \left[\Pi(\alpha|\beta) \neq 0 \Rightarrow \forall
\alpha^\prime \neq \alpha \; \Pi(\alpha^\prime|\beta)=0 \right]
\end{equation}
As $\Pi(\alpha|\beta)=P(\alpha|\beta)$, it follows the first
operation must have been logically deterministic:
\begin{equation}
\forall \beta \; \left[P(\alpha|\beta) \neq 0 \Rightarrow \forall
\alpha^\prime \neq \alpha \; P(\alpha^\prime|\beta)=0 \right]
\end{equation}
Together this means
\begin{equation}
P(\alpha|\beta)=\frac{P(\alpha)}{P(\beta)}=\Pi(\alpha|\beta)=\frac{w_\alpha}{w_\beta}
\end{equation}
and $ \mean{\Delta W^{\prime}_\Pi}+ \mean{\Delta W}=0$, regardless
of the values of $w_\beta$.  This shows, once more, that logically
reversible operations may be thermodynamically optimised without
reference to the probability distribution over their input states.

A corollary to this is worth noting.  While the second logical
operation, if logically reversible, may be implemented and optimised
without reference to the probability distribution over the input
states, its very definition depends upon the probability
distribution over the input states of the first operation.  The
first operation is defined by the set of transition probabilities
$\{P(\beta|\alpha)\}$, while the second is \textit{defined} by
\begin{equation}
\Pi(\alpha|\beta)=P(\alpha|\beta)=\frac{P(\beta|\alpha)P(\alpha)}{\sum_\alpha
P(\beta|\alpha)P(\alpha)}
\end{equation}
There is, in general, only one way to make this independant of
$\{P(\alpha)\}$: if the \textit{first} operation is logically
reversible, then $P(\alpha|\beta) \in \{0,1\}$.  The second
operation is now logically deterministic and $\Pi(\alpha|\beta) \in
\{0,1\}$ does not require the $\{P(\alpha)\}$.

We can summarise this, as follows:  if an operation,
$\{P(\beta|\alpha)\}$ is logically reversible, then it is possible
to calculate a (logically deterministic) reverse operation,
$\{\Pi(\alpha|\beta)\}$, independantly of the first \textit{input}
probability distribution, $\{P(\alpha)\}$.  However, if
$\{P(\beta|\alpha)\}$ is logically indeterministic, then optimising
the reverse operation requires the \textit{output} probability
distribution $\{P(\beta)\}$.

Conversely, if an operation $\{P(\beta|\alpha)\}$ is logically
deterministic, then it is possible to thermodynamically optimise a
(logically reversible) reverse operation $\{\Pi(\alpha|\beta)\}$,
independantly of the first \textit{output} probability distribution
$\{P(\beta)\}$. However, if $\{P(\beta|\alpha)\}$ is logically
irreversible, then the very calculation of the probabilities
$\{\Pi(\alpha|\beta)\}$ require the first \textit{input} probability
distribution $\{P(\alpha)\}$.

In general, it is only for logically deterministic, reversible
operations (which are permutations) that one can construct optimal
reverse operations independantly of the probability distributions.
\subsection{Thermodynamic irreversibility}
\label{sss:thdyncycles} The reverse operations considered in the
preceding discussion have the property of restoring the original
statistical state of the logical system. They do not, in general,
restore the original logical state.  The question of what is the
`correct' thermodynamic entropy to use in such situations is not
uncontroversial and can depend upon differing physical
interpretations of the probabilities of the initial and final
logical states.  It will therefore be helpful to consider an
approach to thermodynamic reversibility which does not depend upon
such definitions.

We will use this to discuss further that the thermodynamic
optimisation of logically irreversible operations is not possible
without specifying the probability distribution over the input
states.  Then we consider two additional sources of thermodynamic
irreversibility that occur in the practical construction of
information processing systems.
\subsubsection{Thermodynamic cycles}
In phenomenological thermodynamics, in any closed cycle, where a
system returns to its initial state, the total heat generated in
heat baths in the process, must satisfy
\begin{equation}
\sum_i \frac{Q_i}{T_i} \geq 0
\end{equation}
As is well known, in statistical mechanics this can no longer be
relied upon.  There is some probability for the equality being
violated.  However, provided the system does return to its initial
macroscopic state with certainty, then
\begin{equation}
\sum_i \frac{\mean{Q_i}}{T_i} \geq 0
\end{equation}
still holds.  We will regard such a cycle for which the equality
holds, to be a thermodynamically reversible cycle, and use the
following definition\footnote{We define the condition in this way to
take into account the fact that for \textit{any} physical process,
it is always trivially possible to find \textit{some} closed cycle
incorporating that process for which inequality is strictly
positive.} of a thermodynamically reversible process:
\begin{quote}
If a given physical process can, in principle, be included in at
least one thermodynamically reversible cycle, then it is a
thermodynamically reversible process.
\end{quote}
To say otherwise would require one either to say that the overall
cycle is thermodynamically reversible, although one of the steps in
the cycle is not (which challenges what it could possibly mean to
refer to that step as thermodynamically irreversible) or to say that
the overall cycle is thermodynamically irreversible, despite the
fact that it restores the original state with certainty and
generates no net heat in any heat bath (and which means that the
entropy of the universe must be the same at the end as the start of
the cycle).

Conversely
\begin{quote}
If a given physical process cannot, even in principle, be included
in \textit{any} thermodynamically reversible cycles, then it is a
thermodynamically irreversible process.
\end{quote}
To avoid interpretational problems over probability, we will require
that the thermodynamically reversible cycle starts, and ends, with
the system in a physical state that represents a fixed logical state
$a$, with certainty.
\subsubsection{Optimal Implementations} Take any logical operation,
defined by the set $\{P(\beta|\alpha)\}$, and construct a physical
implementation of that operation, optimised for the values
$w_\alpha$ and $w_\beta=\sum_\alpha P(\beta|\alpha) w_\alpha$.  This
physical process will implement the $\{P(\beta|\alpha)\}$ operation
regardless of the input state probabilities.

We now also construct two further operations: a logically
reversible, indeterministic operation, generalising the $UFZ$
operation, that acts on $a$ as the sole possible logical input
state, and outputs state $\alpha$ with probability
$P(\alpha)=w_\alpha$; and a logically irreversible, deterministic
operation, generalising $RTZ$, that acts on the logical states
$\{\beta\}$, and always outputs logical state $a$.  The physical
implementation of this second operation is optimised for
probabilities $P(\beta)=w_\beta$.  Both these are well defined
physical processes.

It is clear that the sequence of these three operations forms a
closed cycle, starting and ending in logical state $a$, with
certainty.  It is trivial to show that the optimal implementation of
these operations produces a net thermodynamic cost of zero, over the
course of the cycle.  The cycle is, unquestionably, a
thermodynamically reversible cycle.  The given physical process that
implements the logical operation must, then, be regarded as a
thermodynamically reversible process.
\subsubsection{Suboptimal Implementations} If we had used a
different initial operation, generating the logical state $\alpha$
with probability $P^\prime(\alpha)$, and a final operation optimised
for probabilities $P^\prime(\beta)=\sum_\alpha
P(\beta|\alpha)P^\prime(\alpha)$, then it is straightforward to show
the cost would be
\begin{equation}
W=kT\sum_{\alpha,\beta}P^\prime(\alpha,\beta) \ln
\left[\frac{P^\prime(\alpha)w_\beta}{P^\prime(\beta)w_\alpha}\right]
\geq 0
\end{equation}
Equality is in general reachable if either
$w_\alpha=P^\prime(\alpha)$ or $\forall \alpha,\beta, P(\beta|\alpha)w_\alpha \in (0,w_\beta)$,
the latter being possible only if the logical operation
$\{P(\beta|\alpha)\}$ is logically reversible.

This leaves us with the following conclusions:
\begin{enumerate}
\item For any logical operation $\{P(\beta|\alpha)\}$, there exist
physical implementations of that operation which can be included in
thermodynamically reversible cycles.
\item For any logical operation $\{P(\beta|\alpha)\}$, and for any given probability distribution $P(
\alpha)$ over the input logical states, there exist physical
implementations of that operation which can be included in
thermodynamically reversible cycles.
\item A given physical implementation of logical operation $\{P(\beta|\alpha)\}$ cannot be included in
thermodynamically reversible cycles for generic probability
distributions over the input states unless it is a logically
reversible operation.
\end{enumerate}
It is not possible to characterise a particular physical process,
that implements a logically irreversible operation, as
thermodynamically reversible, independantly of the specification of
the statistical state on which it acts.  Does this mean that we
cannot characterise the physical process as thermodynamically
reversible, at all?

This situation is not unknown in statistical mechanics, or even
phenomenological thermodynamics.  Let us consider a large container,
divided in half by a removable partition, and in the container is a
macroscopic gas.  The pressure on both sides of the partition is
initially equal, and the gas is always kept in isothermal contact
with a single heat bath.  Removing and reinserting the partition is
clearly thermodynamically reversible.

If we slowly, isothermally, slide the partition to the left,
compressing half the gas and expanding the other half, until the
compressed gas occupies only one-third the container, the pressure
on the left side is double the pressure on the right side (net work
is required). Removing and reinserting the partition at this
off-centre position is not thermodynamically reversible.

This thermodynamic irreversibility is not simply due to the
off-centre position of partition.  Start with the partition in the
centre, but now with gas initially prepared to be at twice the
pressure on the right hand side of the container as on the left hand
side.  Simply removing the partition from the centre of the box is
now thermodynamically irreversible.  Isothermally moving the
partition to the left until the left hand side holds only one third
of the container's volume equalises the pressure (and extracts
work).  Now the off-centre removal and reinsertion of the partition
becomes thermodynamically reversible.

The parallel to the model used for logical operations should be
clear\footnote{Indeed, if we are considering a statistical
mechanical N-atom gas, with N=1, it is exactly the same model.}.  A
given sequence of actions cannot, in general, be regarded as
thermodynamically reversible independantly of the state on which
they act.  To describe a phenomenological thermodynamic process as
thermodynamically reversible it is necessary to specify both the
sequence of actions \textit{and the state on which they act} in the
definition of the physical process.  This carries over into
statistical mechanics and, as we have seen above, into the
thermodynamics of computation.

The situation also bears some similarity to data compression from a
signal source. A given coding scheme will only be optimal for a
\textit{particular} distribution of probabilities of signals from
the source.  Should the signals, in fact, be generated with a
different probability distribution, then the mean length of the
encoded signals will be greater than the Shannon information of the
source.  That Shannon's coding theorem is of practical utility
indicates that it is not inconceivable that there may be information
processing problems where the probability distribution over the
logical states may be available when designing optimal physical
implementations.
\subsubsection{Uncertain Operations}
If the logical operation acts upon a set of statistical states, but
it is uncertain which operations have acted upon the system in the
past, an additional source of thermodynamic irreversibility may
occur.  As an example of this, let us consider a bit that has been
deterministically set to either zero or one, from a standard state
$a$, and now needs to be reset to the standard state.

If the first operation set the bit to zero, the operation is
$UFZ(1)$, and the work required was
\begin{equation}
\Delta W_0=(E_0-T_R S_0)-(E_a-T_R S_a)
\end{equation}
and if set to one, $UFZ(0)$ gives
\begin{equation}
\Delta W_1=(E_1-T_R S_1)-(E_a-T_R S_a)
\end{equation}
If the reset operation is optimised with values $w_0+w_1=1$, then it
is $RTZ(w_0)$,
\begin{eqnarray}
\Delta W_{R0}&=&(E_a-T_R S_a)-(E_0-T_R S_0)-kT_R \ln w_0 \nonumber \\
\Delta W_{R1}&=&(E_a-T_R S_a)-(E_1-T_R S_1)-kT_R \ln w_1 \nonumber
\\ &&
\end{eqnarray}
giving total costs
\begin{eqnarray}
\Delta W_{T0}&=&\Delta W_{R0}+\Delta W_0=-kT_R \ln w_0 \geq 0 \nonumber \\
\Delta W_{T1}&=&\Delta W_{R1}+\Delta W_1=-kT_R \ln w_1 \geq 0
\end{eqnarray}
The equalities can be reached by setting $w_0=1$ or $w_1=1$,
respectively, but this is only possible if the other is zero - which
would require an infinite amount of work if the wrong operation had
taken place!

If we assign non-zero probabilities to the set operations of $p_0$
and $p_1$, then the expected cost for the cycle is
\begin{eqnarray}
\Delta W_T&=&-kT_R \sum_{i=0,1} p_i \ln w_i \geq -kT_R \sum_i p_i
\ln p_i
>0 \nonumber \\ &&
\end{eqnarray}
with the equality occurring if $w_i=p_i$.  Clearly this is a
thermodynamically irreversible cycle, despite the fact that each of
the three logical operations ($UFZ(1)$, $UFZ(0)$, $RTZ(w_0)$) can be
individually incorporated in a thermodynamically reversible cycle.
What is the source of the irreversibility?

There are a number of ways one can regard this.  Both the
deterministic set operations are, in themselves, thermodynamically
reversible.  It could be argued that the irreversibility in
whichever of the $\Delta W_{T0}$ or the $\Delta W_{T1}$ cycles that
actually took place, is then through the reset operation, which was
designed for the possibility of either deterministic set operation.

A different way to perceive the situation is to regard the situation
as either being $\Delta W_{T0}$, which may be thermodynamically
optimised by setting $w_0=1$, or $\Delta W_{T1}$ which may be
optimised by $w_1=1$.  In either case the cycle becomes
thermodynamically reversible.  The source of thermodynamic
irreversibility would then be that the reset operation was not
optimised for the correct probabilities (which must now be regarded
as either $p_0=1$ or $p_1=1$, corresponding to which operation
actually did take place).

Yet another way would be to consider a new class of operation: an
`uncertain' operation, where there is an uncertainty as to which
actual operation took place.  In this case we have an `Uncertain
Set' operation, which could be defined as $p_0 \, UFZ(1) +p_1 \,
UFZ(0)$. This operation has a work requirement:
\begin{equation}
\Delta W_U=\sum_{i=0,1}p_i(E_i-T_R S_i)-(E_a-T_R S_a)
\end{equation}
Viewed as a logical operation, this would take as input logical
state $0$ with probability one, and output states $0$ and $1$ with
probabilities $p_0$ and $p_1$.  The optimal implementation of such a
logical transformation of information would be $UFZ(p_0)$, which has
cost
\begin{equation}
\Delta W=\sum_{i=0,1}p_i(E_i-T_R S_i)-(E_a-T_R S_a)+kT_R \sum_i p_i
\ln p_i
\end{equation}
As a logical transformation of information, the `Uncertain Set'
operation is clearly sub-optimal. It is thermodynamically
irreversible, as it cannot be included in any thermodynamically
reversible cycle.

What is the `correct' way to view this?  We are not sure this is a
well-posed question.  However, what all three explanations have in
common is that the thermodynamic irreversibility is a consequence of
the uncertainty over which logical operation took place.  It is this
that prevents the construction of a thermodynamically reversible
cycle.

Suppose we have a number of different process, labeled with
$\gamma$, and each implements a logical operation
$\{P(\beta|\alpha,\gamma)\}$, optimised for input state
probabilities $P(\alpha)$.  The optimal cost for operation $\gamma$
is
\begin{eqnarray}
\Delta W_\gamma&=&\sum_\beta P(\beta|\gamma)\left(E_\beta-T_R
S_\beta +kT_R \ln P(\beta|\gamma)\right) \eqnwrap -\sum_\alpha
P(\alpha)\left(E_\alpha-T_R S_\alpha +kT_R \ln P(\alpha)\right)
\end{eqnarray}
where $P(\beta|\gamma)=\sum_\alpha P(\beta|\alpha,\gamma)P(\alpha)$

We now assign a probability $P(\gamma)$ to each logical operation
occurring (and take for granted
$P(\alpha,\gamma)=P(\alpha)P(\gamma)$).  The cost of this `Generic
Uncertain Operation' is
\begin{eqnarray}
\mean{\Delta W_\gamma}&=&\sum_{\alpha,\beta,\gamma}
P(\alpha,\beta,\gamma)\left(E_\beta-E_\alpha \right. \eqnwrap \left.
 -T_R \left( S_\beta - S_\alpha - k \ln
\frac{P(\beta|\gamma)}{P(\alpha)}\right) \right)
\end{eqnarray}
This produces the output states $\{\beta\}$ with probabilities
$P(\beta)=\sum_{\alpha,\gamma}P(\beta|\alpha,\gamma)P(\alpha)P(\gamma)$.

Now to complete the cycle, we consider an optimised Reset operation,
the acts upon states $\{\beta\}$ to produce the standard state $a$,
and an optimal operation that acts upon $a$ and produces the logical
states $\{\alpha\}$ with probability $P(\alpha)$. Combining these
two has the cost
\begin{eqnarray}
\Delta W_R&=&\sum_{\alpha} P(\alpha)\left(E_\alpha-T_R S_\alpha
+kT_R \ln P(\alpha)\right) \eqnwrap - \sum_{\beta}
P(\beta)\left(E_\beta-T_R S_\beta +kT_R \ln P(\beta)\right)
\end{eqnarray}
giving a total cost for the cycle of
\begin{equation}
\mean{\Delta W_\gamma}+\Delta W_R= kT_R \sum_{\beta,\gamma}
P(\beta,\gamma)\ln \frac{P(\beta,\gamma)}{P(\beta)P(\gamma)}\geq 0
\end{equation}
Equality is reached only if $P(\beta,\gamma)=P(\beta)P(\gamma)$,
i.e. there is no correlation between the occurrence of the $\beta$
output states and which $\gamma$ operation actually took place.  The
thermodynamic irreversibility that occurs if $P(\beta|\gamma)\neq
P(\beta)$ does not depend upon whether the operation required to
restore the original statistical state is logically reversible or
logically irreversible.

In the familiar case of the `Uncertain Set'-Reset cycle there is a
compression of the logical state space during the reset operation
and the compensating increase in the non-information bearing degrees
of freedom of system or environment may give the impression that the
source of the thermodynamic irreversibility is the logical
irreversibility of the Reset operation.  The `Generic Uncertain
Operation' shows this is not the case.  In fact an optimal operation
that restores the $P(\alpha)$ distribution from the $P(\beta)$
distribution could be logically reversible and the cycle still be
thermodynamically irreversible provided $P(\beta|\gamma)\neq
P(\beta)$.  It is the uncertainty over which $\gamma$ operation took
place that is the source of the thermodynamic irreversibility.

As before, this situation has well known parallels in standard
statistical mechanics. The spread of gas molecules into a box,
shielded from any outside interference, can in principle be
reversed.  (Spin-echo experiments have even demonstrated similar
reversals to this in the laboratory.) However, this reversal is very
sensitive to uncertainty in the outside forces that act upon the
gas.  In a famous calculation, Borel showed that the gravitational
influence of remote stars could change the microscopic state of an
expanding macroscopic gas within seconds.  Reversing that expansion
would be possible, in principle, if there was highly detailed
knowledge of the gravitational influence of the remote bodies on the
gas (or if microscopic state of the expanded gas molecules turned
out to be independant of that influence) but becomes impossible when
the gravitational influence is uncertain.

\subsubsection{Partial Operations}
A third reason for the occurrence of thermodynamic irreversibility
is that the physical implementation of the logical operation is not
able to take into account the existence of correlations between
systems, and can only act upon part of the total logical
state\footnote{See also \cite{Andersen2008}.}.  We will show that,
in this case, logically reversible operations are able to avoid the
thermodynamic irreversibility, although logically irreversible
operations are still not \textit{always} thermodynamically
irreversible.

Suppose the input logical states factorise into the product of two
subsystems, with the logical states of the first system in the set
$\{\alpha\}$ and the second system in $\{\gamma\}$, so the joint
system is described by the logical states $\{(\alpha,\gamma)\}$. Now
consider a logical operation that acts only on the $\alpha$ states,
with probabilities $P(\beta|\alpha)$.  If the physical
implementation of this logical operation has no access to the
$\gamma$ system, then the physical implementation can only be
optimised with respect to the marginal probabilities
\begin{equation}
P(\alpha)=\sum_\gamma P(\alpha,\gamma)
\end{equation}
The system ends up in output states from the product of the states
of the $\{\gamma\}$ and $\{\beta\}$ systems, $\{(\beta,\gamma)\}$,
with probabilities
\begin{equation}
P(\beta,\gamma)=\sum_\alpha P(\beta|\alpha)P(\alpha,\gamma)
\end{equation}

The resulting thermodynamic cost of the partially optimised
operation is:
\begin{eqnarray}
\Delta W_P&=& \sum_{\beta} P(\beta)\left(E_\beta-T_R \left(S_\beta
-k \ln P(\beta)\right)\right) \eqnwrap -\sum_{\alpha}
P(\alpha)\left(E_\alpha-T_R \left(S_\alpha -k \ln
P(\alpha)\right)\right)
\end{eqnarray}
where $P(\beta)=\sum_\gamma P(\beta,\gamma)$ and we have assumed
\begin{eqnarray}
E_{\alpha,\gamma}&=&E_\alpha+E_\gamma \nonumber \\
E_{\beta,\gamma}&=&E_\beta+E_\gamma \nonumber \\
S_{\alpha,\gamma}&=&S_\alpha+S_\gamma \nonumber \\
S_{\beta,\gamma}&=&S_\beta+S_\gamma
\end{eqnarray}

An optimal operation for restoring the states $(\alpha, \gamma)$,
with probabilities $P(\alpha,\gamma)$ has a thermodynamic cost of
\begin{eqnarray}
\Delta W_R&=&\sum_{\alpha,\gamma} P(\alpha,\gamma)\left(E_\alpha-T_R
\left(S_\alpha-k \ln P(\alpha,\gamma)\right)\right)\eqnwrap -
\sum_{\beta,\gamma}P(\beta,\gamma)\left(E_\beta-T_R \left(S_\beta
-k\ln P(\beta,\gamma)\right)\right)
\end{eqnarray}
so the net cost for the cycle is
\begin{eqnarray}
\frac{\Delta W_R+\Delta W_P}{kT_R}& = & kT_R \left(
\sum_{\alpha,\gamma} P(\alpha,\gamma)\ln
\frac{P(\alpha,\gamma)}{P(\alpha)} \right. \eqnwrap
-\left.\sum_{\beta,\gamma}P(\beta,\gamma)\ln
\frac{P(\beta,\gamma)}{P(\beta)}\right)
\end{eqnarray}
This can be expressed as changes in conditional or correlation
information:
\begin{eqnarray}
\frac{\Delta W_R+\Delta W_P}{kT_R}&=& - \sum_{\alpha,\beta,\gamma}
P(\alpha,\beta,\gamma)\left(
 \ln P(\gamma|\beta)-\ln P(\gamma|\alpha)
 \right) \nonumber \\
&=& -\sum_{\alpha,\beta,\gamma} P(\alpha,\beta,\gamma)\left(
 \ln \frac{P(\beta,\gamma)}{P(\beta)P(\gamma)} \right. \eqnwrap \left.-\ln
 \frac{P(\alpha,\gamma)}{P(\alpha)P(\gamma)}
\right)
\end{eqnarray}
Using the identity
\begin{equation}
P(\alpha,\gamma|\beta)P(\beta|\alpha)=P(\beta,\gamma|\alpha)P(\alpha|\beta)
\end{equation}
gives the form of conditional correlations:
\begin{eqnarray}
\frac{\Delta W_R+\Delta W_P}{kT_R}&=& - \sum_{\alpha,\beta,\gamma}
P(\alpha,\beta,\gamma)\left(
 \ln \frac{P(\beta,\gamma|\alpha)}{P(\beta|\alpha)P(\gamma|\alpha)} \right. \eqnwrap
 \left. -\ln
 \frac{P(\alpha,\gamma|\beta)}{P(\alpha|\beta)P(\gamma|\beta)}
\right)
\end{eqnarray}
As $P(\beta|\alpha,\gamma)=P(\beta|\alpha)$, then
\begin{eqnarray}
P(\gamma,\beta|\alpha) & = & P(\beta|\alpha,\gamma)P(\gamma|\alpha)
\nonumber \\
 &=&P(\beta|\alpha)P(\gamma|\alpha)
\end{eqnarray}
The $\alpha$ states screen off any correlation between the $\beta$
and $\gamma$ states and the first term is zero, so:
\begin{equation}\label{eq:partialcorr}
\frac{\Delta W_R+\Delta W_P}{kT_R}=\sum_{\alpha,\beta,\gamma}
P(\alpha,\beta,\gamma)\ln
\frac{P(\alpha,\gamma|\beta)}{P(\alpha|\beta)P(\gamma|\beta)} \geq 0
\end{equation}
Equality occurs if, and only if, $\beta$ screens off any
correlations between $\alpha$ and $\gamma$:
\begin{equation}\label{eq:betascreens}
P(\alpha,\gamma|\beta)=P(\alpha|\beta)P(\gamma|\beta)
\end{equation}

This can happen directly if there is no initial correlation between
the $\alpha$ and $\gamma$ systems, so that
$P(\alpha,\gamma)=P(\alpha)P(\gamma)$. With
$P(\beta|\alpha,\gamma)=P(\beta|\alpha)$ it follows
$P(\alpha,\beta,\gamma)=P(\alpha,\beta)P(\gamma)$ and from that
Equation \ref{eq:betascreens} holds, as might be expected.

To see the effect of logical reversibility, rewrite Equation
\ref{eq:partialcorr} as
\begin{equation}
\frac{\Delta W_R+\Delta W_P}{kT_R}=\sum_{\alpha,\beta,\gamma}
P(\alpha,\beta,\gamma)\left( \ln P(\alpha|\gamma,\beta) -\ln
P(\alpha|\beta) \right)
\end{equation}
If the operation is logically reversible $P(\alpha|\beta)\in
\{0,1\}$.  This gives
\begin{eqnarray}
P(\alpha|\beta)=0 & \Rightarrow &  P(\alpha,\beta,\gamma)=0 \nonumber \\
P(\alpha|\beta)=1 & \Rightarrow &  P(\alpha|\gamma,\beta)=1
\end{eqnarray}
and the summation is identically zero.  Logically reversible
operations avoid the thermodynamically irreversible
cost\footnote{However, one should note that it is still possible for
some logically \textit{irreversible} operation to satisfy the
conditions for thermodynamic reversibility, Equation
\ref{eq:betascreens}, for particular correlations between the
$\alpha$ and $\gamma$ systems.}.

\section{Conclusions}\label{s:conclusion}
The focus on the process of Landauer {\em Erasure} can give the
impression that Landauer's {\em principle} should be exclusively
about the thermodynamics of logically irreversible processes and
further that the heat generation of such processes implies
thermodynamic irreversibility:
\begin{quote}
To erase a bit of information in an environment at temperature $T$
requires dissipation of energy $\geq kT \ln 2$. \cite{Cav90,Cav93}

in erasing one bit \ldots of information one dissipates, on average,
at least $k_B T \ln\left(2\right)$ of energy into the environment.
\cite{Pie00}

a logically irreversible operation must be implemented by a
physically irreversible device, which dissipates heat into the
environment \cite{Bub02}

erasure of one bit of information increases the entropy of the
environment by at least $k \ln 2$ \cite{LR03}[pg 27]

any logically irreversible manipulation of data $\ldots$ must be
accompanied by a corresponding entropy increase in the
non-information bearing degrees of freedom of the information
processing apparatus or its environment.  Conversely, it is
generally accepted that any logically reversible transformation of
information can in principle be accomplished by an appropriate
physical mechanism operating in a thermodynamically reversible
fashion. \cite{Ben03}
\end{quote}

though it should be noted that not all advocates of Landauer's
principle regard the process of erasure as {\em necessarily}
thermodynamically irreversible:

\begin{quote}
a logically irreversible operation \ldots may be thermodynamically
reversible or not depending on the data to which it is applied.  If
it is applied to random data \ldots it is thermodynamically
reversible, because it decreases the entropy of the data while
increasing the entropy of the environment by the same amount
\cite{Ben03}
\end{quote}

In \cite{Mar05b} it was argued that there exists a valid
thermodynamically reverse process to Landauer Erasure, but which
needs to be classified as logically indeterministic,
  which
we called Reverse Landauer Erasure (or $RLE$).  Consideration of the
thermodynamic consequences of the existence of this process led us
to conclude there was no convincing evidence that logically
irreversible operations had special thermodynamic characteristics.
Instead, we hypothesised that a generalised form of Landauer's
principle should be possible that made no reference to
irreversibility, whether logical or thermodynamic. This was
expressed in two conjectures:

\begin{quote}\textbf{(E)}:
{\em Any} logically irreversible transformation of information can
in principle be accomplished by an appropriate physical mechanism
operating in a thermodynamically reversible fashion.

\textbf{(F)}: A logical operation needs to generate heat equal to at
least $-k T \ln 2$ times the change in the total quantity of Shannon
information over the operation, or:
\[
\Delta W \geq kT \ln 2 [ H_i-H_f ]
\]
\raggedleft \cite{Mar05b}[pg. 362]
\end{quote}

In this paper we have both proved and generalised these conjectures.
Our approach has been to take the widest definition of logical
operations available and most general procedure for physically
implementing these operations that we can. This requires us to
consider logically indeterministic operations as well as
deterministic ones, logically reversible operations as well as
irreversible ones.

Other papers have made some consideration of Landauer Erasure in the
context of non-uniform temperatures\cite{Sch94},
entropy\cite{Shi95,Fah96}, and energy\cite{Pie00}, while
\cite{Barkeshli2005} combines varying entropy and energy.
Non-uniform input probabilities are considered in the proofs of
\cite{Pie00,Shi95}. The thermodynamics of logically indeterministic
operations does not seem to be considered before
\cite{Mar02,Mar05b}, although \cite{Pen70}[Chapter VI] is close, and
it is noticeable that \cite{Ben03} refers throughout to
deterministic computation.  During the preparation of this paper, a
paper has appeared by Turgut\cite{Turgut2006} deriving similar
results using classical phase space arguments.

General proofs of Landauer's principle seem hard to come by (as
pointed out in \cite{Nor05}) although \cite{Pie00} derives similar
results to those of Section \ref{ss:nobetter}, but restricted to the
$RTZ$ operation, and under the assumption that logical states are
represented by pure quantum states (an assumption shared with
\cite{Jac05}).  Here we allow logical states to be represented by
density matrices and consider any logical operation.  We have
considered the most general setting for physically implementing
classical logical operations, covering and extending these earlier
results. We derive the most general statement of Landauer's
principle, prove it cannot be exceeded and give a limiting process
which can achieve it.

The general statement of Landauer's principle we arrived at is:

\begin{quote}
\textbf{Generalised Landauer's principle}

A physical implementation of a logical transformation of information
has minimal expectation value of the work requirement given by:
\begin{equation}
\mean{\Delta W} \geq \mean{\Delta E} - T\Delta S
\end{equation}
where $\mean{\Delta E}$ is the change in the mean internal energy of
the information processing system,  $\Delta S$ the change in the
Gibbs-von Neumann entropy of that system and $T$ is the temperature
of the heat bath into which any heat is absorbed.

The equality is reachable, in principle, by any logical operation,
and if the equality is reached the physical implementation is
thermodynamically reversible.
\end{quote}

We have then shown how various additional assumptions and
simplifications can lead to more familiar versions of Landauer's
principle that can be found in the literature and these are special
cases of the GLP.  Generalisations about the relationship between
information processing and thermodynamic entropy based upon these
special cases can be misleading.

In particular, we have argued, counter to a widespread version of
Landauer's principle, that there is nothing in principle, that
prevents a logically irreversible operation from being implemented
in a thermodynamically reversible manner.  What differs between
logically irreversible operations and logically reversible
operations is that to thermodynamically optimise physical
implementations of the former it is necessary to take into account
the probability distribution over the complete set of input logical
states. A physical implementation of a logically irreversible
operation, optimised for a particular input probability
distribution, will not be thermodynamically irreversible for a
different input probability distribution.  If the physical
implementation cannot access a correlated system, then logically
irreversible operations may incur additional costs.

As the practical business of actually building physical devices to
implement logical operations will typically not be able to make such
optimisations, it is natural to assume an equiprobable distribution
over a subsystem, and expect thermodynamic irreversibility.
Nevertheless the point remains: \textit{in principle} it is always
possible to physically implement logically irreversible
transformations of information in thermodynamically reversible ways.
There are many practical reasons why a logically irreversible
operation may not be thermodynamically optimised, and it is clearly
important and useful to explore such problems.  In this paper,
however, we are primarily concerned with the question: what is the
\textit{fundamental limit} for thermodynamically optimising the
physical implementation of a given logical operation?

We have demonstrated that, under the same conditions of uniform
computing that imply logically deterministic, irreversible
operations generate heat, logically indeterministic, reversible
operations extract heat from the environment which can be converted
into work.  At the same time we have demonstrated that under other
conditions, adiabatic equilibrium computing, information processing
is able to progress without any exchange of work or heat, regardless
of the type of logical operation.

The thermodynamic reversibility of all logical operations is, of
course, based upon the definition of thermodynamic reversibility
given in Sections \ref{ss:gvnentropy} and \ref{sss:thdyncycles}.
Other approaches to thermodynamics (such as
\cite{Shenker2000,Nor05,SLGP05}) use different concepts of entropy
and correspondingly different definitions of thermodynamic
irreversibility to this paper.  Ultimately the most important
question is not what particular quantity one chooses to label as
`thermodynamic' entropy.    The GLP we have derived here is valid,
whether one chooses to regard the Gibbs-von Neumann entropy as the
true `thermodynamic' entropy, or not.  What is important is the
actual work required to drive a system, the actual heat generated by
that system.  As there is no disagreement over the fundamental
microscopic dynamics, it would be surprising if we were unable to be
able to agree on these values, regardless of the definition of
entropy to which we choose to adhere.

\textbf{Acknowledgements} The author would like to thank Basil
Hiley, Keith Bowden, Chris Timpson, John Barrett, James Ladyman and
Tony Short for various discussions on subjects related to
computation, information and entropy and Steve Weinstein for his
comments and advice.  The author would also like to thank an
anonymous referee for raising the question of the `Uncertain Set'
operation.

Research at Perimeter Institute for Theoretical Physics is supported
in part by the Government of Canada through NSERC and by the
Province of Ontario through MRI.

\begin{appendix}
\section{One bit logical operations}\label{ap:onebit}
\subsection{Do Nothing: $IDN$}
The simplest operation is
\begin{eqnarray}
P(\beta=0|\alpha=0)&=&1 \nonumber \\
 P(\beta=0|\alpha=1)&=&0  \nonumber \\
P(\beta=1|\alpha=0)&=&0  \nonumber \\
P(\beta=1|\alpha=1)&=&1
\end{eqnarray}
giving
\begin{eqnarray}
P(\alpha=0|\beta=0)&=&1  \nonumber \\
P(\alpha=0|\beta=1)&=&0  \nonumber \\
P(\alpha=1|\beta=0)&=&0  \nonumber \\
P(\alpha=1|\beta=1)&=&1
\end{eqnarray}
and is logically deterministic and reversible.

\subsection{Logical NOT: $NOT$} Logical NOT, acting upon an input
bit with probability $p$ of being in state $0$, is very simple:

\begin{eqnarray}
P(\beta=0|\alpha=0)&=&0  \nonumber \\
P(\beta=0|\alpha=1)&=&1 \nonumber \\
P(\beta=1|\alpha=0)&=&1  \nonumber \\
P(\beta=1|\alpha=1)&=&0
\end{eqnarray}
giving
\begin{eqnarray}
P(\alpha=0|\beta=0)&=&0  \nonumber \\
P(\alpha=0|\beta=1)&=&1  \nonumber \\
P(\alpha=1|\beta=0)&=&1  \nonumber \\
P(\alpha=1|\beta=1)&=&0
\end{eqnarray}
This is logically deterministic and reversible.

\subsection{Reset To Zero: $RTZ(p)$} If the input state $0$ occurs
with probability $p$, then the $RTZ(p)$ operation has the
properties:

\begin{eqnarray}
P(\beta=0|\alpha=0)&=&1  \nonumber \\
P(\beta=0|\alpha=1)&=&1
\end{eqnarray}
giving
\begin{eqnarray}
P(\alpha=0|\beta=0)&=&p  \nonumber \\
P(\alpha=1|\beta=0)&=&1-p 
\end{eqnarray}
This is logically deterministic and irreversible.  As $\forall
\alpha \ P(\beta=1|\alpha)=0$ the state $\beta=1$ is not an output
state of the operation and we leave it out of the table.

\subsection{Unset From Zero: $UFZ(p)$} The reverse operation to
$RTZ$, where the state $0$ is taken to state $0$ with probability
$p$, will be called here the UNSET FROM ZERO operation.  In
\cite{Mar05b} this operation was described in terms of the physical
process that reverses $LE$, so was
 called `Reverse Landauer Erasure' or $RLE$.  In this paper we will refer to the logical
 operation as $UFZ$, and to the specific physical process that can be used to embody it as $RLE$.
 This operation may also be characterised as a random number generator.

\begin{eqnarray}
P(\beta=0|\alpha=0)&=&p  \nonumber \\
P(\beta=1|\alpha=0)&=&1-p 
\end{eqnarray}
giving
\begin{eqnarray}
P(\alpha=0|\beta=0)&=&1  \nonumber \\
P(\alpha=0|\beta=1)&=&1 
\end{eqnarray}
This is indeterministic but reversible. As $\forall \beta \
P(\alpha=1|\beta)=0$ the state $\alpha=1$ is not an input state of
the operation and we leave it out of the table.

\subsection{Randomize: $RND(p,p^{\prime})$} The operation which
takes an input probability of $p$ of the state being $0$ and
produces $0$ with an output probability of $p^{\prime}$, regardless
of input state:

\begin{eqnarray}
P(\beta=0|\alpha=0)&=&p^{\prime}  \nonumber \\
P(\beta=0|\alpha=1)&=&p^{\prime} \nonumber \\
P(\beta=1|\alpha=0)&=&1-p^{\prime}  \nonumber \\
P(\beta=1|\alpha=1)&=&1-p^{\prime}
\end{eqnarray}
giving
\begin{eqnarray}
P(\alpha=0|\beta=0)&=&p  \nonumber \\
P(\alpha=0|\beta=1)&=&p  \nonumber \\
P(\alpha=1|\beta=0)&=&1-p  \nonumber \\
P(\alpha=1|\beta=1)&=&1-p
\end{eqnarray}
This is indeterministic and irreversible.

We note that $RTZ(p) \equiv RND(p,1)$ and $UFZ(p) \equiv RND(1,p)$.

\subsection{General One Bit: $GOB(p,p_{00},p_{11}$)}
Finally, we consider the most generic operation possible for 1 input
bit and 1 output bit. The operation can be wholly defined by one
input probability $p$ and two conditional probabilities $p_{00}$ and
$p_{11}$
\begin{eqnarray}
P(\alpha=0)&=&p \nonumber \\
P(\alpha=1)&=&1-p \nonumber \\
P(\beta=0|\alpha=0)&=&p_{00} \nonumber \\
P(\beta=0|\alpha=1)&=&1-p_{11} \nonumber \\
P(\beta=1|\alpha=0)&=&1-p_{00} \nonumber \\
P(\beta=1|\alpha=1)&=&p_{11}
\end{eqnarray}
giving
\begin{eqnarray}
P(\alpha=0,\beta=0)&=&p p_{00}  \nonumber \\
P(\alpha=0,\beta=1)&=&p (1-p_{00})  \nonumber \\
P(\alpha=1,\beta=0)&=&(1-p) (1-p_{11})  \nonumber \\
P(\alpha=1,\beta=1)&=&(1-p) p_{11}
\end{eqnarray}
and
\begin{eqnarray}
P(\beta=0)&=&p p_{00} + (1-p) (1-p_{11}) \nonumber \\
P(\beta=1)&=&p (1-p_{00})+(1-p) p_{11}
\end{eqnarray}
so
\begin{eqnarray}
P(\alpha=0|\beta=0)&=&\frac{p p_{00}}{p p_{00} + (1-p) (1-p_{11})}  \nonumber \\
P(\alpha=0|\beta=1)&=&\frac{p (1-p_{00})}{p (1-p_{00})+(1-p) p_{11}}  \nonumber \\
P(\alpha=1|\beta=0)&=&\frac{(1-p) (1-p_{11})}{p p_{00} + (1-p)(1- p_{11})}  \nonumber \\
P(\alpha=1|\beta=1)&=&\frac{(1-p) p_{11}}{p (1-p_{00})+(1-p)p_{11}}
\end{eqnarray}
In general, this is logically indeterministic and irreversible, but
can become logically reversible or deterministic under the right
limits:
\begin{eqnarray}
IDN &\equiv& GOB(p,1,1)  \nonumber \\
NOT &\equiv& GOB(p,0,0)  \nonumber \\
RTZ(p) &\equiv& GOB(p,1,0)  \nonumber \\
UFZ(p) &\equiv& GOB(1,p,-)  \nonumber \\
RND(p,p^{\prime}) &\equiv& GOB(p,p^{\prime},1-p^{\prime})
\end{eqnarray}
\end{appendix}

\end{document}